\documentclass[twocolumn, trackchanges]{aastex631}
\usepackage{wrapfig}

\begin{document}

\title{Identification of New Candidate Be/X-Ray Binary Systems in the Small Magellanic Cloud via Analysis of S-CUBED Source Catalog}

\author[0009-0000-6957-8466]{Thomas Martin Gaudin}
\affiliation{Pennsylvania State University}

\author[0000-0002-6745-4790]{Jamie Kennea}
\affiliation{Pennsylvania State University}

\author[0000-0002-0763-8547]{M.J. Coe}
\affiliation{Physics and Astronomy, The University of Southampton, SO17 1BJ, UK}

\author{Phil A. Evans}
\affiliation{Leicester University, UK}

% \author{Jamie Kennea}
% \affiliation{Pennsylvania State University}



\begin{abstract}

It has long been known that a large population of Be/X-ray Binaries (BeXRBs) exists in the Milky Way’s neighboring dwarf galaxy, the Small Magellanic Cloud (SMC), due to a recent period of intense star formation. Since 2016, efforts have been made to monitor this population and identify new BeXRBs through the Swift SMC Survey (S-CUBED). S-CUBED’s weekly observation cadence has identified many new BeXRBs that exist within the SMC, but evidence suggests that more systems exist that have thusfar escaped detection. A major challenge in identifying new BeXRBs is their transient nature at high-energy wavelengths, which prevents them from being detected via their X-ray emission characteristics when not in outburst. In order to identify sources that may have been missed due to a long period of quiescence, it becomes necessary to devise methods of detection that rely on wavelengths at which BeXRBs are more persistent emitters. In this work, we attempt to use archival analysis of infrared, optical, and ultraviolet observations to identify new candidate BeXRBs that have been overlooked within the S-CUBED source catalog. Using X-ray/optical selection of source properties, unsupervised clustering, SED-fitting to VizieR archival measurements, and ultraviolet light curve analysis, we are able to identify six new candidate BeXRB systems that otherwise would have been missed by automated analysis pipelines. Using these results, we demonstrate the use of ultraviolet through near-infrared observational data in identifying candidate BeXRBs when they cannot be identified using their X-ray emission.

\end{abstract}



\section{Introduction} \label{sec:intro}

% Overall feeling about the paper: A very good start. Needs more figures! Also please redo figures as PDF or postscript so they don't look so awful.

Be/X-ray Binaries (BeXRBs) are a common type of High Mass X-ray Binary (HMXB) that are characterized by a main sequence OB star being orbited by a compact object (see \citealt{2011Reig} for a review). Typically the compact object is a neutron star (NS), but white dwarfs \citep{2020Coe, 2021Kennea} have also been observed. The OB star derives its ``Be" designation from the Balmer-series Hydrogen emission lines that are present in its optical spectrum. It is also observed to have an excess of infrared (IR) radiation compared to a normal OB star. These two observational properties are indicators of a geometrically thin, circumstellar ``decretion" disk that surrounds the Be star \citep{2003Porter}. 

At high energies, BeXRBs are transient sources of emission. Interactions between the NS and the Be disk can produce one of two different types of X-ray outburst \citep{1986Stella}, and it is only during these outbursts that BeXRBs can be reliably observed. Type I X-ray outbursts typically reach luminosites of $L_X \sim 10^{36}-10^{37}$ erg s$^{-1}$ \citep{2001Okazaki}, and Type II outbursts reach larger ($L_X \gtrsim 10^{37}$ erg s$^{-1}$), sometimes super-Eddington luminosities \citep{2017Townsend}. However, the fraction of time spend in a state of outburst is typically small \citep{2018Kennea}. These systems spend long periods of time in a quiescent state in between X-ray outbursts. While in quiescence, the X-ray luminosity ($L_X \sim 10^{32} - 10^{35}$ erg s$^{-1}$) of a BeXRB system is below the detection threshold of all but the most powerful X-ray telescopes (see studies by \citealt{2007Rutlidge, 2013Rothschild},  and \citealt{2020Rouco}).

BeXRBs are complex systems that have long been studied, but many open questions still remain. Open questions include the effect of the NS on the formation and structure of the Be disk \citep{2013Rivinius} and the specific mechanisms by which accretion occurs in a system with a highly-magnetized NS \citep{2007Becker}. The drive to improve our understanding of BeXRBs and answer these remaining open questions has led to an active effort to identify and confirm the existence of new systems. The Small Magellanic Cloud (SMC), a dwarf galaxy of the Milky Way, has long been a target of searches for new BeXRBs. X-ray surveys of the SMC have consistently been undertaken by many observatories such as ASCA \citep{2000Yokogawa}, ROSAT \citep{1996Kahabka, 2000Haberl}, XMM-Newton \citep{2013Sturm}, and Chandra \citep{2003Zezas, 2009Antoniou}. Surveys such as these have uncovered an abnormally large population of HMXBs \citep{2015Coe, 2017Yang}, and almost all of these are BeXRBs. This large, nearly homogeneous population of BeXRBs is most recently estimated to contain 111 sources \citep{2025Treiber}. Such a large population is thought to be the result of both a recent period of increased star formation \citep{2004Harris, 2014Rezaeikh} within the dwarf galaxy as well as the lower metallicity environment contributing to the higher rotational velocity of massive stars \citep{2007Martayan}. It makes the SMC the perfect target for continued X-ray surveys that aim to identify new BeXRB candidates and monitor their evolution over time. 

Since 2016, the largest effort to survey the SMC for new BeXRBs has been undertaken by the Swift SMC Survey (S-CUBED) \citep{2018Kennea}. This survey utilizes the X-ray Telescope (XRT; \citealt{2005Burrows}) and UV/Optical Telescope (UVOT; \citealt{2005Roming}) onboard the \textit{Neil Gehrels Swift Observatory} to perform a survey of the SMC in the ultraviolet (UV) and X-ray regimes. S-CUBED performs weekly observations of the SMC in an effort to find new BeXRBs and monitor the X-ray emission of known systems. The weekly cadence of S-CUBED allows for frequent monitoring of known transient or persistent SMC high-energy sources to identify any recent changes to their X-ray luminosity. It also allows for the detection of new X-ray transients that may have appeared within the SMC since the last observation. This survey design is particularly effective for identifying transient BeXRB outbursts. Within the first year of the survey, S-CUBED had identified over 30 BeXRBs \citep{2018Kennea} in the SMC and more confirmed sources have been added in subsequent years \citep{2020Kennea, 2020Monageng, 2021Coe}. 

Despite many years of effort to complete the BeXRB census of the SMC, new sources are still regularly identified (e.g. SXP 341.8; \citealt{2023Maitra}). These newly-identified sources provide strong evidence that there is still a hidden population of BeXRBs that have not yet been identified within the SMC. The difficulty of detecting these hidden sources is likely due to their transient nature in the X-ray regime. BeXRBs are typically identified via their X-ray outbursts, but this is inefficient due to the relative infrequency of these events. In order to properly characterize the BeXRB population of any galaxy, it thus becomes necessary to develop methods of identifying candidate BeXRB systems in the near-infrared (NIR) through UV regimes where the main sequence companion star is expected to be a persistent, bright source. 

The typical optical signature used to identify a BeXRB is the presence of the variable, double-peaked H$\alpha$ line embedded in the spectrum of an OB-type main sequence star \citep{2011Reig}. However, BeXRBs have been known to go through periods of disk loss \citep{2021CoeB} during which no emission lines are observed. Additionally, spectroscopic monitoring is a process that requires many high-resolution ground-based observations to complete. It would be far more efficient if photometric data could be used instead so that large populations can be characterized swiftly and archival observations of candidates can be utilized effectively. It then follows that the use of a multi-wavelength spectral energy distribution (SED) and light curve data could greatly speed up the identification of BeXRB candidates via their optical counterparts. 

There have been studies that have used broadband photometry to identify Be stars, but the methods employed have been far different from those outlined in this work. Typical methods have involved the use of a combination of broadband UV/optical and narrowband H$\alpha$ photometry to identify color excesses typical of Be stars by way of differential photometry and color-magnitude diagrams \citep{2018Milone, 2024Navarete}. To the best of our knowledge, the combined use of SED-fitting and UV variability represents a novel method for the identification of BeXRBs in the Magellanic Clouds. 

In this work, we report the results of an archival analysis of S-CUBED data. Our archival analysis identifies six new candidate BeXRB systems that have not yet been detected in the SMC via X-ray survey. Using the novel combination of IR to UV SED-fitting of archival photometric data and 
UVOT light curve analysis, we report evidence of optical companions to six transient sources that were first identified by S-CUBED but have limited X-ray data available for analysis. The rest of the paper will be organized as follows. Section \ref{sec:observations} will describe the S-CUBED survey in more detail. Section \ref{sec:methods} will describe the methods used for this identification and confirmation procedure. Section \ref{sec:results} will present the results of our search efforts and describe the data available for all candidate sources. Section \ref{sec:discussion} will discuss the reasons by which we can confidently state that these six candidate systems are indeed BeXRBs and report on the detection of a new X-ray outburst from one of our candidate systems, Swift J010902.6-723710. 

\section{S-CUBED Observations} \label{sec:observations}

The S-CUBED survey has been ongoing since 2016 with the first year results being published by \citet{2018Kennea}. The weekly cadence allows for the emission of known BeXRBs to be monitored at a frequency that is greater than their expected orbital period ($P_{orb} \sim 10 - 100$ days). S-CUBED \textbf{observes} 149 overlapping tiles that are selected to \textbf{ensure} continuous coverage of the entire dwarf galaxy. The survey's 142 tiles are designed to account for the pointing uncertainties inherent to Swift by including a 3-arcminute overlap between tiles that is well-within the circular (radius of 11.8 arcminutes) field of view for XRT \citep{2018Kennea}. However, due to the much smaller, square-shaped field of view of UVOT (17x17 arcminutes; \citealt{2005Roming}), there are gaps between UVOT fields that do not get monitored by S-CUBED. These gaps are do not cover fixed locations on the sky. Instead, they vary due to pointing uncertainty and the roll angle that is optimal for the spacecraft to maintain star-tracker lock and effectively regulate temperature. A 50-60s snapshot is obtained for each tile at both X-ray and UV wavelengths. Each snapshot is taken with XRT in photon counting (PC) mode and UVOT observing with its \textit{uvw1}-band filter, centered at 2600\AA.

All XRT data taken as part of the S-CUBED survey is automatically processed using an X-ray data analysis pipeline created by the UK Swift Data Science Center \citep{2009Evans, 2014Evans}. The automatic pipeline will flag any quiescent sources that have recently entered an outburst phase for further monitoring. It also identifies sources of X-ray emission that repeatedly appear in the same location and produces crucial data products for each source. For each source, an observation-binned 0.3-10 keV-band light curve is produced. If a source is detected during an S-CUBED observation, then the XRT count rate is reported from which the flux and luminosity of the source can be derived. Otherwise, an upper limit for the count rate is reported when the source is not detected. Additionally, the cumulative 0.3-10 keV-band X-ray spectrum is reported for each source. This spectrum is automatically fit to an absorbed power law model using \texttt{xspec} \citep{Arnaud96}, and the best-fitting values for column density along the line of sight ($N_H$) and photon index ($\Gamma$) are reported for any sources that can be fitted. Some sources have been detected at faint X-ray luminosities or few observations and thus, do not have a well-defined X-ray spectra. In these cases, no spectral fit can be obtained and the best-fitting spectral parameters of such systems remain an open question.

Each observation contributes to the cumulative X-ray data available for analysis of a given system, leading to deeper observations and improved results as the survey progresses. A quality flag of ``Good", ``Reasonable", or ``Poor" is assigned to each detected source, which relates to the likelihood that it is a real astrophysical source. \citet{2018Kennea} and \citet{2020Evans} discuss these quality flags in more detail and explain the false detection probability that each flag corresponds to. UVOT data is not automatically analyzed for each source, so manual data processing is required in order to retrieve \textit{uvw1}-band photometric information. More detailed information about the reduction process is provided in Subsection \ref{subsec: UVOT}.

S-CUBED has seen great success in identifying X-ray sources within the SMC. After the first year of the survey, \citet{2018Kennea} published a catalogue of over 700 X-ray sources that had been detected as part of the survey. Seven years later, over 2000 sources have been detected by S-CUBED, and many of these sources remain un-identified. There are known BeXRBs and other sources of X-ray emission such as supernova remnants within the S-CUBED catalog, but there are also many sources that still cannot be identified. It is likely that many unidentified sources are background active galactic nuclei (AGN) which are known to be common around the SMC \citep{2024Ivanov}. However, some of the unidentified sources are likely quiescent BeXRBs that have spent the duration of the survey in a quiescent state. 

Automated analysis of S-CUBED data is performed each week for individual observations of S-CUBED data, therefore the pipeline utilized is optimized for transient detection. Only sources that have shown evidence of a transient X-ray outburst in the most recent observation are flagged for follow-up observations. Due to the long periods of quiescence that can be experienced by BeXRB systems, it is likely that some systems have not produced a transient X-ray outburst at a luminosity that can be reliably detected by S-CUBED. Instead, these sources would only be detected by Swift when their quiescent luminosity was high and would be missed by our automated pipeline. Only through analysis of the cumulative X-ray data available for each source can a quiescent BeXRB be detected. The methods outlined in this paper represent the first efforts to select for these potential hidden candidate BeXRBs using cumulative S-CUBED data products. 

\section{Methods} \label{sec:methods}

\subsection{Archival Data Mining} \label{subsec:mining}

In order to determine which population of sources may represent candidate BeXRBs, we first performed an archival search within the S-CUBED source catalog. In order to be 99.7\% confident that any candidate is a real astrophysical source \citep{2020Evans}, we only kept sources in our sample that are flagged as ``Good". The data was also filtered by the photon index derived from the X-ray spectrum of each source. BeXRBs are expected to have a hard X-ray spectrum with a photon index that is approximately 1. In order to account for standard errors in power law fits to the X-ray spectrum, only sources for which the confidence interval of the photon index intersected with the range of photon indices between 0.5 and 1.5 were kept. All others were removed. Many sources, particularly those with only two or three detections by XRT have no spectral fits, so any source with no photon index was included as well. 

Once a sample of systems had been generated based on the above criteria, those corresponding to known sources of X-ray emission were removed. A query was performed for each candidate using the Set of Identifications, Measurements and Bibliography for Astronomical Data (SIMBAD) \citep{2000Wenger} database. This SIMBAD search was used to locate known sources of X-ray emission within a 15 arcsecond radius of the center of the XRT error region. For the purposes of this query, a known source of radiation could be an AGN, a protostar/young stellar object, a supernova remnant, or a known/candidate BeXRB. If any known sources were detected, then it was assumed that the nearby known object was the source of the XRT detection. All candidates that could be matched to a known source of X-ray emission were removed. At the end of SIMBAD filtration, the only sources left in the sample were unknown sources with a photon index between 0.5 and 1.5 or no spectral fit. 

One last archival search was then carried out to remove any sources without a candidate Be star in close proximity. For SMC BeXRBs, all systems are expected to have an optical companion that is bright in the $U$ and $B$-bands ($>16$ magnitude in the $B$-band). The presence of circumstellar disk material in companion Be stars is additionally expected to produce a positive $I-R$ magnitude \citep{2011Reig}. In order to determine whether a star matching these criteria could be located nearby, a query was performed using the VizieR \citep{2000Ochsenbein} database aggregator. Given the small size of a typical XRT error region ($\sim 5$ arcseconds), the query was limited to stars within 10 arcseconds of the XRT source location. There are many catalogs that give stellar magnitude information, so to narrow down results, we limit our search to results from the \textit{Gaia} mission \citep{2016GaiaCollaboration}, the Optical Gravitational Lensing Experiment (OGLE; \citealt{1997Udalski}), the Naval Observatory Merged Astrometric Dataset (NOMAD; \citealt{2004Zacharias}), the Whole-Sky USNO-B1.0 catalog \citep{2003Monet}, and the Guide Star Catalog (GSC; \citealt{2008Lasker}). Any candidate with no nearby stars matching our criteria was removed from the sample.

By using this multi-stage archival data search, we were able to identify 20 S-CUBED sources that were likely to be a hidden quiescent BeXRB. These sources were then analyzed by the methods described in Section \ref{subsec:sed} and Section \ref{subsec: UVOT} to determine if there was any merit to their inclusion on this list of candidate BeXRBs.

\subsection{Extinction Correction} \label{subsec:extinction}

Before performing SED-fitting to our modified blackbody curve, a wavelength-dependent correction must be applied to account for interstellar extinction. Extinction is caused by dust that is present along the line-of-sight towards a source \citep{1990Mathis}. Dust along the path can be located both in the Milky Way and in the SMC, and its composition can vary depending on the location of a source within the dwarf galaxy \citep{2020Gorski}. Accurate extinction correction must account for the amount of reddening that is present along the specific line-of-sight towards each source, because each path will be slightly different. 

In order to perform extinction corrections, the reddening map of \citet{2021Skowron} was utilized. This is a detailed 3.4-arcminute resolution reddening map based on OGLE observations of red clump stars. The reddening along each line-of-sight is reported as a median $E(V-I)$ value that accounts for the effects of both foreground Milky Way and internal SMC dust. A conversion factor between $E(V-I)$ and $E(B-V)$ is presented in Equation 11 of \citet{2021Skowron}:
\begin{equation}
    E(V-I) = 1.237 \, E(B-V)
\end{equation}
This equation is originally used to convert between OGLE $E(V-I)$ values and the reddening map $E(B-V)$ values generated by \citet{1998Schlegel}. The conversion factor of 1.237 is derived using Table 6 of \citet{2011Schlafly}, which re-calibrates the map of Schlegel. 

For each source, the $E(V-I)$ value is retrieved by searching for the closest line-of-sight position. Once this value is determined, it is then converted to $E(B-V)$. \citet{2011Schlafly} recommends the use of the Fitzpatrick \citep{1999Fitzpatrick} extinction law with $R_V = 3.1$. The $E(B-V)$ values determined for each source can be used to calibrate the Fitzpatrick extinction law and determine the total extinction, $A_{\lambda}$, that needed to be corrected for each data point in the SED. Given the total extinction at each wavelength, the unreddened flux for each photometric measurement is calculated using the following equation:
\begin{equation}
    f_{\lambda_{unred}} = 10^{0.4 A_\lambda} f_{\lambda}
\end{equation}
where $f_\lambda$ is the initial extincted flux density value of each data point. 

\begin{figure}
    \centering
    \includegraphics[scale=0.33]{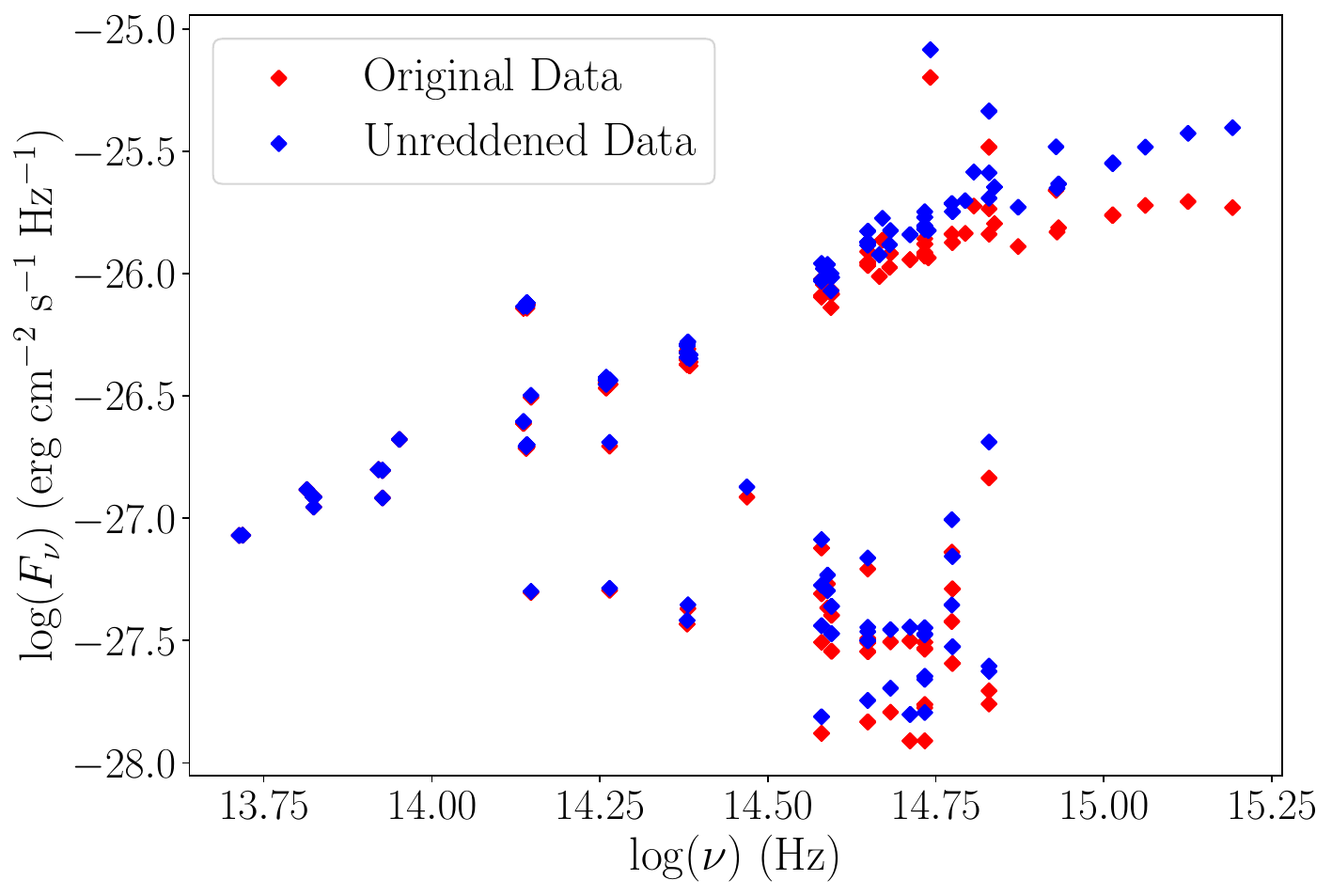}
    \caption{Example of the results of extinction corrections applied to a source in the SMC. Red points represent the original data downloaded from the VizieR photometry database. Blue points represent the flux measurements after extinction correction has been applied. The flux increases by almost half an order of magnitude for points in the optical/near-UV wavelength regimes, implying that the temperature and radius of candidate OB stars would be underestimated without applying this correction.}
    \label{fig:reddening}
\end{figure}

The results of applying extinction correction to all photometry for a source in the SMC are shown in Figure \ref{fig:reddening}. In this figure, all photometric measurements made for stars in the vicinity of Swift J010902.6-723710 have had extinction corrections applied to them. The importance of applying extinction correction is apparent, particularly for OB stars. Optical and near-UV photons are the most affected by dust grain extinction along the line-of-sight. As this is the regime where their stellar blackbody curve peaks, the radius and temperature of an SMC OB stars are likely to be significantly underestimated without applying these corrections.  

\subsection{SED Fitting} \label{subsec:sed}

The magnitudes of nearby stars are not sufficient evidence to match an optical counterpart to an S-CUBED X-ray source. It is possible that a foreground star may have a similar magnitude to an SMC OB main sequence star, or that there may be an un-related bright OB star or giant star near the position of an S-CUBED source. Additionally, the SMC is inherently a crowded field, so there may be more than one bright star nearby to an S-CUBED source which creates confusion when trying to identify potential optical counterparts. In order to better identify optical counterparts, all available nearby photometric data must be included in the analysis. A spectral energy distribution (SED) was generated for each source using all available archival IR-UV data. Fundamental stellar parameters were then derived for all nearby stars using SED-fitting techniques. Because the SMC has a well-constrained distance, a distance-modified blackbody curve can be used to determine the stellar radius and effective temperature for a star in the SMC. The modified flux density for an SMC star is given by:
\begin{equation}
    F(\nu, R, T) = \pi \left( \frac{R}{D} \right)^2 \left( \frac{2 h \nu^3}{c^2} \frac{1}{e^{\frac{h \nu}{k T}}-1}\right) 
\end{equation}
where $D$ is the accepted distance to the SMC of 62.44 kpc \citep{2020Graczyk}, $R$ is the stellar radius, and $T$ is the effective stellar temperature. Radius and Temperature are the free parameters that can be constrained using SED-fitting. Once the stellar radius and effective temperature are known, enough information is available to determine the stellar luminosity. The combination of these three parameters is enough to place constraints on the spectral type and luminosity class of the star in a manner that is both computationally and observationally inexpensive. There are other methods that may be able to improve the uncertainties that are derived for the fundamental stellar parameters. However, we have chosen a method that allows us to quickly identify likely candidates even if it means accepting larger error bars.

\begin{figure}
    \centering
    \includegraphics[scale=0.17]{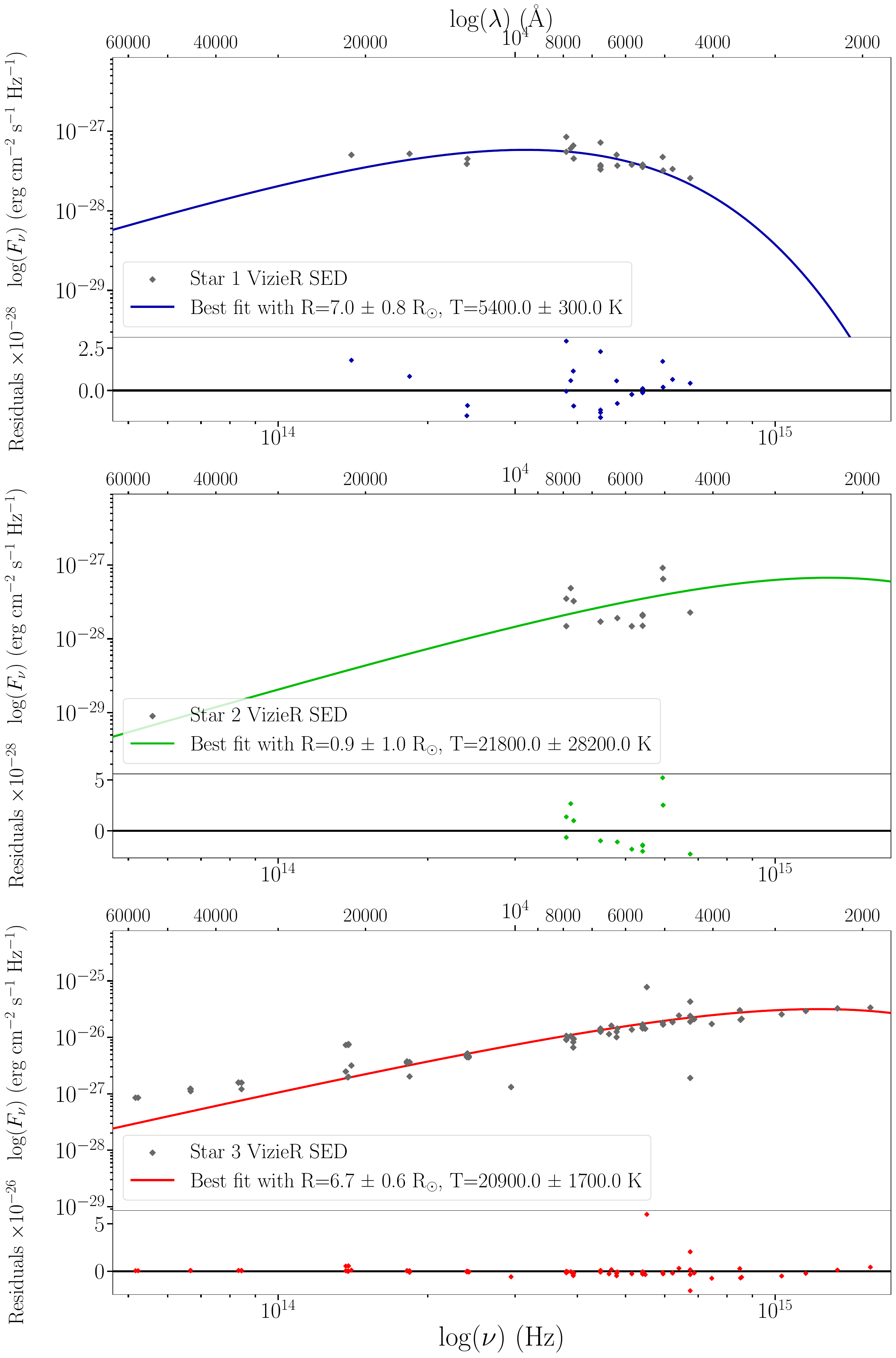}
    \caption{Example of best-fitting modified Planck function for each star within 8 arcseconds of the S-CUBED target. Data for each star was retrieved from the VizieR photometry database. Nearby stars were identified using the \textit{Gaia} database.}
    \label{fig:1772fitting}
\end{figure}

The SED data used in SED-fitting was extracted using the VizieR photometry tool \citep{2000Ochsenbein}, which allows for all cataloged flux measurements found within a given radius of a coordinate position to be extracted for analysis. For each candidate source, the \textit{Gaia} Data Release 3 database was queried to determine the positions of all stars within 8 arcseconds of the XRT centroid position. The radius of 8 arcseconds was chosen to ensure that all stars that could feasibly overlap with the XRT error region for an S-CUBED source (typically $\sim$5 arcseconds in radius) would be included in the analysis. For each star, an IR-UV SED was generated for all measurements made within a 1.0 arcsecond radius of the stellar position given by \textit{Gaia}. The small 1.0 arcsecond radius is chosen in order to limit the effect of spurious data points that have been assigned to the wrong star due to the aforementioned crowded field. 

In order to prevent duplicate stars from being fit to the above SED model multiple times, a filter had to be applied to identify and remove potential duplicate stars. Duplication removal was accomplished by calculating the separation between the reported \textit{Gaia} position of each pair of stars within the field. If a pair of stars was found to be separated by less than an arcsecond, the stars were assumed to be duplicates, and one of the two was removed from the list. Only after all duplicates were identified and removed could the SED data be downloaded for each star. For some sources, there were no duplicates to remove. More commonly, one or two duplicates were found and removed. No source in the sample had more than 5 duplicates that had to be removed before we could proceed.

Once the data were downloaded, all flux measurements brighter than 12th magnitude (0.06 Jy) were filtered out as they were likely to correspond to foreground stars instead of SMC stars. This matches observational data for the brightest optical companions to known S-CUBED sources \citep{2018Kennea, 2015Coe} which have an upper limit magnitude of $\sim$12. After applying an extinction correction (see Section \ref{subsec:extinction} for more details) to all data, a method of non-linear least squares curve fitting from the Lmfit Python package \citep{matt_newville_2024_10998841} was then applied to each star in order to determine the values for stellar temperature and radius that produce the best fit. For approximately half of the measurements included in the VizieR photometry tool, no error is given for the reported flux measurement. Instead of removing half of the SED points available for fitting, we elected to perform curve-fitting without the inclusion of flux measurement errors.

An example of the resulting fits from this process can be seen in Figure \ref{fig:1772fitting}, where the blackbody fitting process outlined above has been applied to Swift J010902.6-723710. In this case, there were three nearby stars identified by \textit{Gaia}.  For all sources identified by \textit{Gaia}, the best-fitting parameter values for stellar radius and effective temperature were determined to be those which minimized the sum of the squared residuals for each star. The errors are estimated by taking the square of the covariance matrix produced by the fitting results. Two of the three stars fit best to fundamental stellar parameters that are realistic for Hydrogen-burning stars, so the blackbody curve with the best-fitting stellar parameters is plotted alongside the stellar SED. The best-fitting stellar parameters derived for the third source (``Star 2") are not consistent with those of any known type of star. Therefore, we can say that it is likely not a candidate OB star in close proximity to our X-ray source. It is possible that a faint background AGN with a flat optical SED is being mis-identified by \textit{Gaia}, but fitting an AGN SED to the available photometric data is beyond the scope of this paper. 

\subsection{Light Curve Analysis} \label{subsec: UVOT}

\begin{figure*}
    \centering
    \includegraphics[scale=0.59]{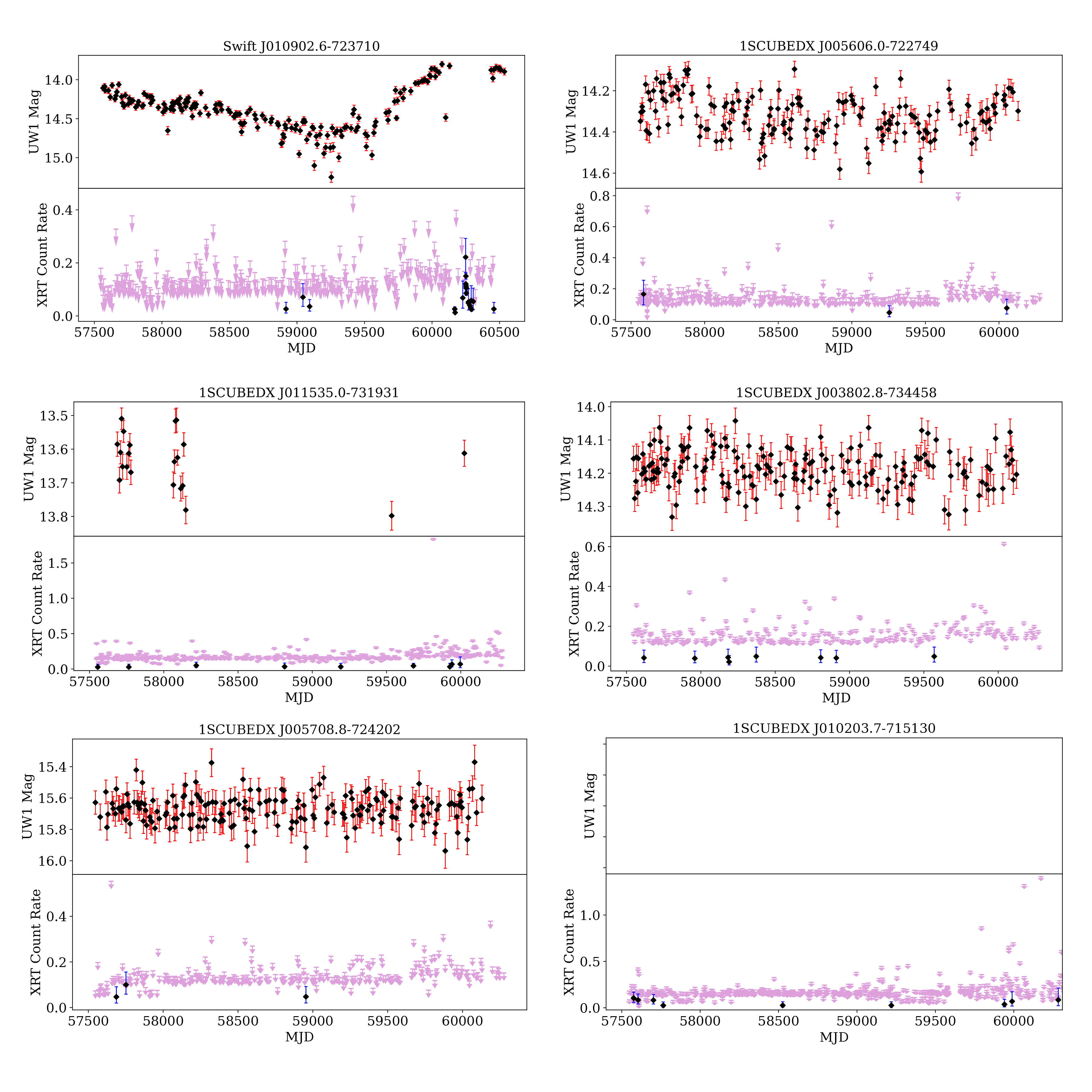}
    \caption{Combined UVOT and XRT light curves for 5 of the 6 candidate BeXRBs identified via archival analysis. For all light curves, the \textit{uvw1}-band magnitude is plotted above the XRT count rate for the source. Arrows represent upper limits for the XRT flux when the source was not detected. No UVOT light curve is available for 1SCUBEDX J010203.7-715130 as its position on the sky places it in a gap between UVOT tiles.}
    \label{fig:all lcs}
\end{figure*}

One of the most prominent features of a BeXRB system in both the optical and IR bands is the presence of strong variability. Multi-year light curves produced by OGLE have long been used to study these systems, particularly improving our understanding of the Be star and its surrounding disk material \citep{2015Coe, 2025Treiber}. OGLE data have been used to demonstrate variability on both short timescales that correspond to plausible orbital periods of the system ($\sim$10-1000 days) and much longer timescales that can sometimes mimic super-orbital periodic behavior.

Somewhat surprisingly, the multi-year \textit{uvw1}-band light curves that are generated using weekly S-CUBED observations have consistently been found to duplicate the variability results of OGLE. UV variability has been observed in BeXRBs that contain both NS \citep{2020Kennea, 2021CoeB, 2024Coe} and WD compact objects \citep{2020Coe, 2021Kennea, 2024Gaudin}. Additionally, the UV variability of a BeXRB system is often found to be correlated with the \textit{I}-band variability of the system. The results of S-CUBED suggest that the presence of UV variability is a fundamental feature of BeXRB systems.

With these results in mind, we attempt to identify any persistent UV emission in the vicinity of SMC X-ray sources and search for variability. We expect that the presence of a massive OB star will manifest itself in UVOT observations as a persistent source in the \textit{uvw1}-band. Additionally, we expect the multi-year \textit{uvw1}-band light curve to show signs of variability if the source is a BeXRB. If one of the 20 sources in the sample shows no evidence of persistent UV emission or the UV emission does not vary on either a plausible orbital period or super-orbital period timescale, then it is likely that the source is not a true BeXRB. Thus, all sources that show no signs of UV variability are excluded from the final candidates list.  

S-CUBED UVOT data was extracted for each source using the Swift Target of Opportunity (TOO) Application Programming Interface (API). Using the API, all UVOT S-CUBED frames were retrieved within a distance of 17 arcminutes to the XRT position of each source. A light curve was then generated using the tools provided by the FTOOLS software package \citep{1999Blackburn} that is maintained by NASA's High Energy Astrophysics Science Archive Research Center (HEASARC)\footnote{FTOOLS Webpage: \href{http://heasarc.gsfc.nasa.gov/ftools}{http://heasarc.gsfc.nasa.gov/ftools}}. FTOOLS contains Swift-specific functions that are used for the analysis of UVOT data. Before extracting photometry for each source, frames from the first year of the survey needed to be aspect corrected so that all tiles have the correct angular coordinates assigned to each pixel. This is accomplished using the \texttt{uvotunicorr} method in FTOOLS. Once each frame has the proper aspect correction, then \texttt{uvotsource} was used to extract UVOT photometry from each frame for a 5 arcsecond radius around the XRT source position. In order to ensure that the counts within the source radius are properly calibrated, a nearby background region containing no stars was created with an 8 arcsecond radius. Figure \ref{fig:all lcs} shows the light curves generated using this method.

Once a light curve was generated for each source, it was then analyzed for evidence of persistent, bright UV emission. We define persistent UV emission to be any source with average \textit{uvw1}-band magnitude brighter than 17. If there was sufficient evidence that a source was persistent in the UV, then it was checked for variability. Based on results from previous studies such as \citet{2020Kennea} and \citet{2021Coe}, the UV variability on the timescale of weeks is expected to be on the order of 0.1 magnitudes. On longer timescales ($\sim$500-1000 days), the variability can be much greater, sometimes varying by almost a full magnitude in brightness, but there are also cases \citep{2021Coe} where no long timescale variability is observed. A source must vary over at least short timescales for it to be considered to be a candidate BeXRB.

\begin{table*}[t]
    \begin{center}
     \begin{tabular}{c|c|c|c|c|c|c|c|c}
        SC\# & 1SCUBEDX & $\alpha \, (^\circ)$ & $\delta \, (^\circ)$ & Spectral Fit? & $E(V-I)$ & Sep (") & $T_{opt}$ (K) & $R_{opt}$ (R$_{\odot}$)  \\
        \hline\hline
        SC72 & J005606.0-722749 &  14.02518 & -72.4639 & N & 0.054 &  1.81 & $22600 \pm 1400$ & $5.4 \pm 0.3$ \\
        SC131 & J010203.7-715130 &  15.51550 & -71.8585 & N & 0.053 & 3.65 & $14300 \pm 1400$ & $4.8 \pm 0.6$ \\
        SC151 & J011535.0-731931 & 18.89583 & -73.3254 & N & 0.157 & 3.89 & $36900 \pm 3200$  & $2.3 \pm 0.2$ \\
        SC251 & J003802.8-734458 & 9.51183 & -73.7495 & Y & 0.041 & 3.94 & $10500 \pm 200$ & $17.0 \pm 0.5$ \\
        SC430 & J005708.8-724202 &14.28670 & -72.7007 & N & 0.083 & 2.42 & $14800 \pm 3300$ & $5.1 \pm 1.3$\\
        SC1772 & - & 17.26102 & -72.6197 & Y & 0.089 & 1.69 & $20900 \pm 1700$ & $6.7 \pm 0.6$ \\
    \end{tabular}
    \end{center}
    \caption{Table containing all derived parameters that provide evidence in favor of the BeXRB nature of the 6 candidate sources identified via archival analysis. The first two columns identify each candidate using the standard nomenclature first adopted by \citet{2018Kennea} to name sources that are identified by S-CUBED. The columns headed by $\alpha$ and $\delta$ correspond to the decimal degree Right Ascension and Declination positions of each object on the sky. The next two columns represent X-ray properties of the sources. The Spectral Fit column identifies whether enough XRT detections have occurred for the 0.3-10 keV spectrum to be fit by an absorbed power law. The $E(V-I)$ column reports the line-of-sight extinction value that is derived for each source from the extinction map of \citet{2021Skowron}. Finally, the last three columns report the results of SED-fitting for the best optical counterpart candidate. ``Sep" refers to the separation in arcseconds between the \textit{Gaia} position and the center of the XRT error region. $T_{opt}$ and $R_{opt}$ are the stellar temperature and radius that achieve the best fit to our modified blackbody model.}
    \label{tab:properties}
\end{table*}

The Swift UVOT field-of-view is 17'x17', which is smaller than the XRT field-of-view (11.8' radius circle). Because of this size difference in the fields-of-view, it is possible for a source to be observable by overlapping XRT tiles, but not by any UVOT tiles. This is the case for two sources, Swift J010203.7-715130 and Swift J011535.0-731931. Both sources are included in the final list of candidates, but a UVOT light curve cannot be generated for Swift J010203.7-715130. It has been omitted from Figure \ref{fig:all lcs}. A light curve has been generated for Swift J011535.0-731931 and included in Figure \ref{fig:all lcs}, but it is relatively sparse as the source is not often visible to UVOT.

\section{Results} \label{sec:results}

Using the methods outlined in Section \ref{sec:methods}, the list of 2014 total SMC X-ray sources was filtered down to 20 sources of interest based on the Archival Data Mining method of Section \ref{subsec:mining}. A further 9 sources were removed via the methods of SED Fitting presented in Section \ref{subsec:sed}. The Light Curve Analysis method presented in Section \ref{subsec: UVOT} was used to remove another 5 sources. At the end of this processes, 6 candidate BeXRBs are identified based on their IR, optical, and UV properties. These candidates and their properties are listed in Table \ref{tab:properties}. There were many sources that were investigated but did not make this final candidates list because they failed one or more of the checks above. The most common check that was failed by those that did not make the list is the SED fitting check performed using the methods of Section \ref{subsec:sed}. All of these sources are revealed to be in close proximity to no stars that match the fundamental stellar parameters expected for an OB star in the Small Magellanic Cloud. Other sources were found to not be UV-variable, which limits their probability to be a BeXRB system based on historical trends associated with S-CUBED observations.

Based on the derived properties of each source that are presented in the table, there are reasons to suspect that each of the 6 sources that pass all of the checks outlined in Section \ref{sec:methods} are indeed BeXRBs. Below, the argument for binarity in each source is outlined in more detail.

\subsection{Swift J010902.6-723710 (SC1772)}\label{subsec:Source1772}

\begin{figure}
    \centering
    \includegraphics[scale=0.25]{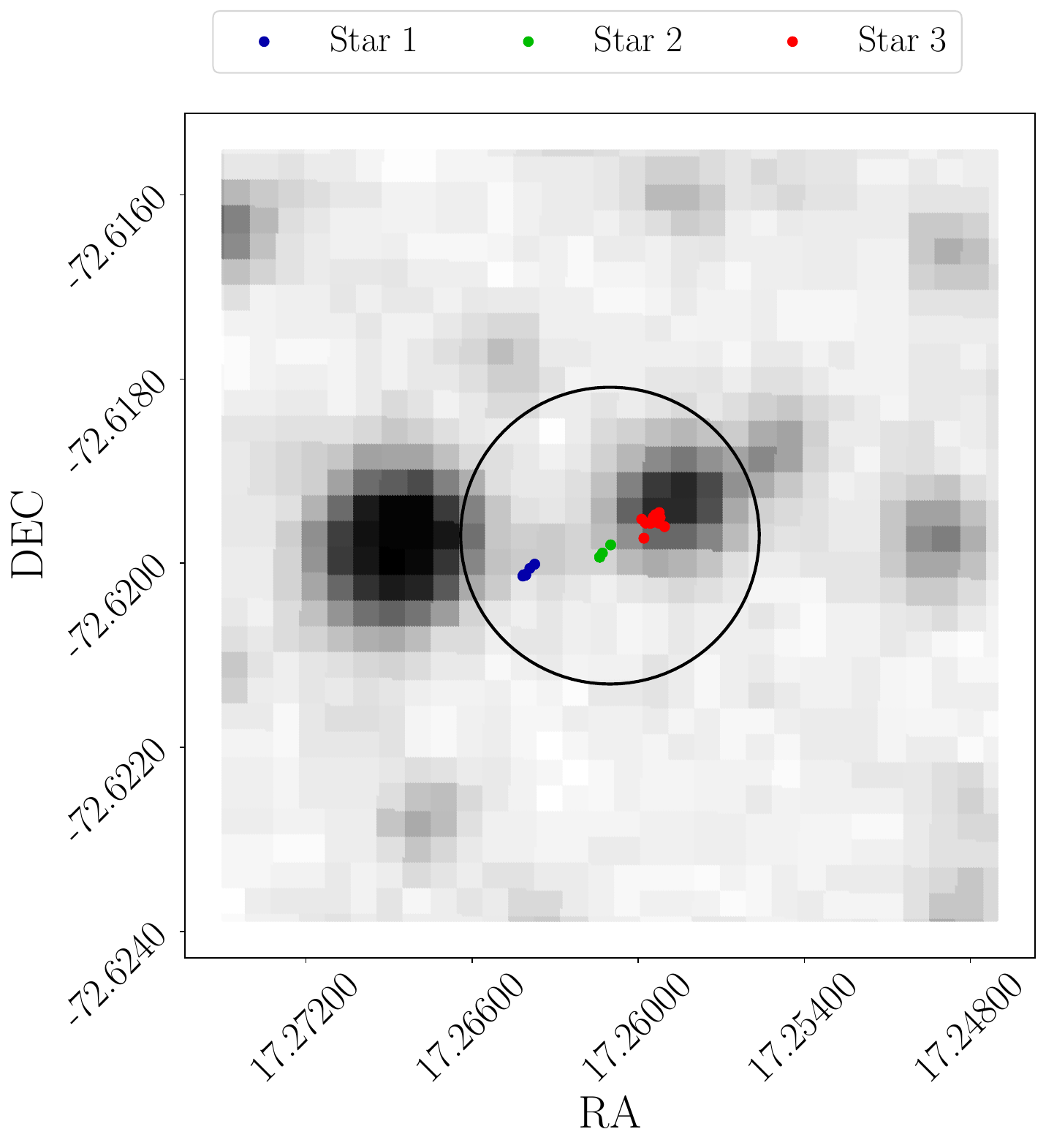}
    \caption{The locations of all photometric data for each star detected by \textit{Gaia} near Swift J010902.6-723710. Photometric positions are plotted over a Digitized Sky Survey (DSS; \citealt{2020STScI}) reference image for the region. The black circle represents the Swift XRT error region for the source. Only three stars are found in close proximity to this source. }
    \label{fig:1772clustering}
\end{figure}

The methods outlined above first identified Swift J010902.6-723710 as a candidate BeXRB in May of 2023 when work on this project first began. At this point in time, no transient X-ray outburst had ever been detected for the source, so it could not be confirmed as a new BeXRB system. Despite a lack of transient emission, there had been multiple bright detections of the source that allowed for a detailed X-ray spectrum to be produced. This is one of only two sources in the sample that has produced a detection by XRT with enough significance for the automatic pipeline outlined in Section \ref{sec:observations} to produce a fit to the 0.3-10 keV X-ray spectrum of the object. An absorbed power law fit to the XRT spectrum using \texttt{xspec} indicates that Swift J010902.6-723710 has a hard X-ray spectrum with a photon index of $0.56_{-0.16}^{+0.15}$, which is within the expected range for a BeXRB system. As seen in Figure \ref{fig:1772clustering}, there are three stars nearby to the Swift XRT error region that can be identified using \textit{Gaia}. The SED-fitting shown in Figure \ref{fig:1772fitting} reveals that Star 3 is the best candidate optical counterpart, producing a likely temperature of $20900 \pm 1700$ K and a stellar radius of $6.7 \pm 0.6$ R$_{\odot}$. The \textit{Gaia} position of this star indicates that it has a separation of 1.69 arcseconds to the XRT position of Swift J010902.6-723710. This position is also coincident with a known star, OGLE J010902.25-723710.1, which was flagged as a Be star candidate by \citet{2002Mennickent}. All of these combined factors make Star 3 a strong candidate for the optical counterpart of the X-ray source.

Analysis of the UVOT light curve from Figure \ref{fig:all lcs} reveals that Swift J010902.6-723710 is a persistently bright object in the \textit{uvw1}-band. The average magnitude for the duration of S-CUBED is 14.4, but the UV magnitude is variable both on short and long timescales. S-CUBED monitoring began in April 2016 when the source brightness was measured to be 14.1 magnitude. For approximately the next 2000 days, the source dimmed appreciably, reaching an average of 14.7 magnitude by October 2021. At this point, the source began a period of sustained brightening that lasted until October 2023 when it reached a maximum \textit{uvw1}-band magnitude of 13.73. Similar \textit{uvw1}-band brightening behavior has been observed in the lead up to outbursts from SMC BeXRBs in the past \citep{2020Kennea}. On October 9th, 2023, an outburst was indeed detected by Swift, confirming this source as a BeXRB and providing validation that the methods outlined above can be successful at detecting overlooked BeXRB systems. A more detailed analysis of this outburst is the subject of a previously-published manuscript \citep{2024Gaudin}. 

\subsection{1SCUBEDX J005606.0-722749}\label{subsec:Source72}

\begin{figure}
    \centering
    \includegraphics[scale=0.25]{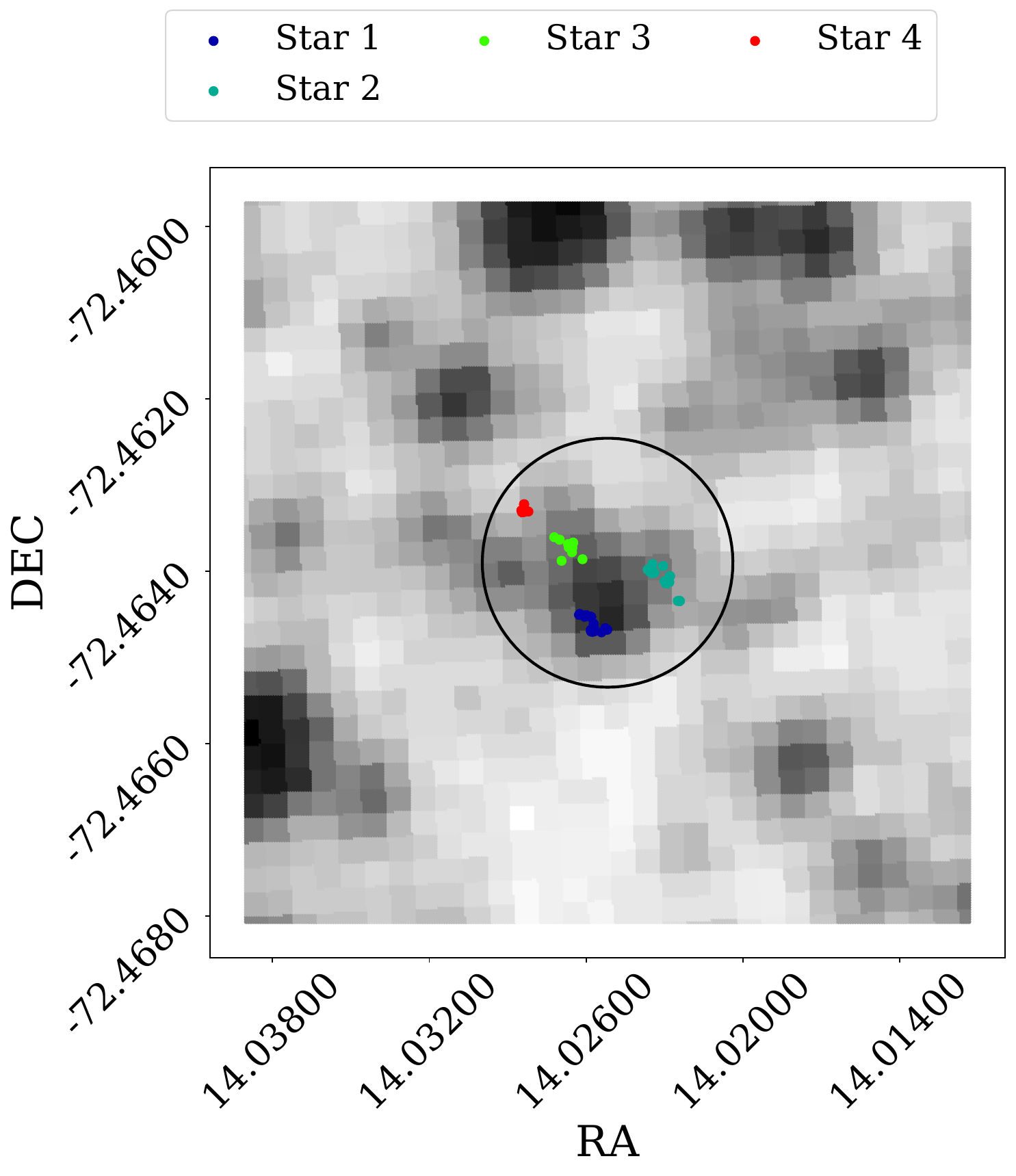}
    \caption{The locations of all photometric data for each star detected by \textit{Gaia} near 1SCUBEDX J005606.0-722749. Photometric positions are plotted over a DSS \citep{2020STScI} reference image for the region. The black circle represents the Swift XRT error region for the source. Nine stars are found in close proximity to this source.}
    \label{fig:72clustering}
\end{figure}

\begin{figure*}
    \centering
    \includegraphics[scale=0.2]{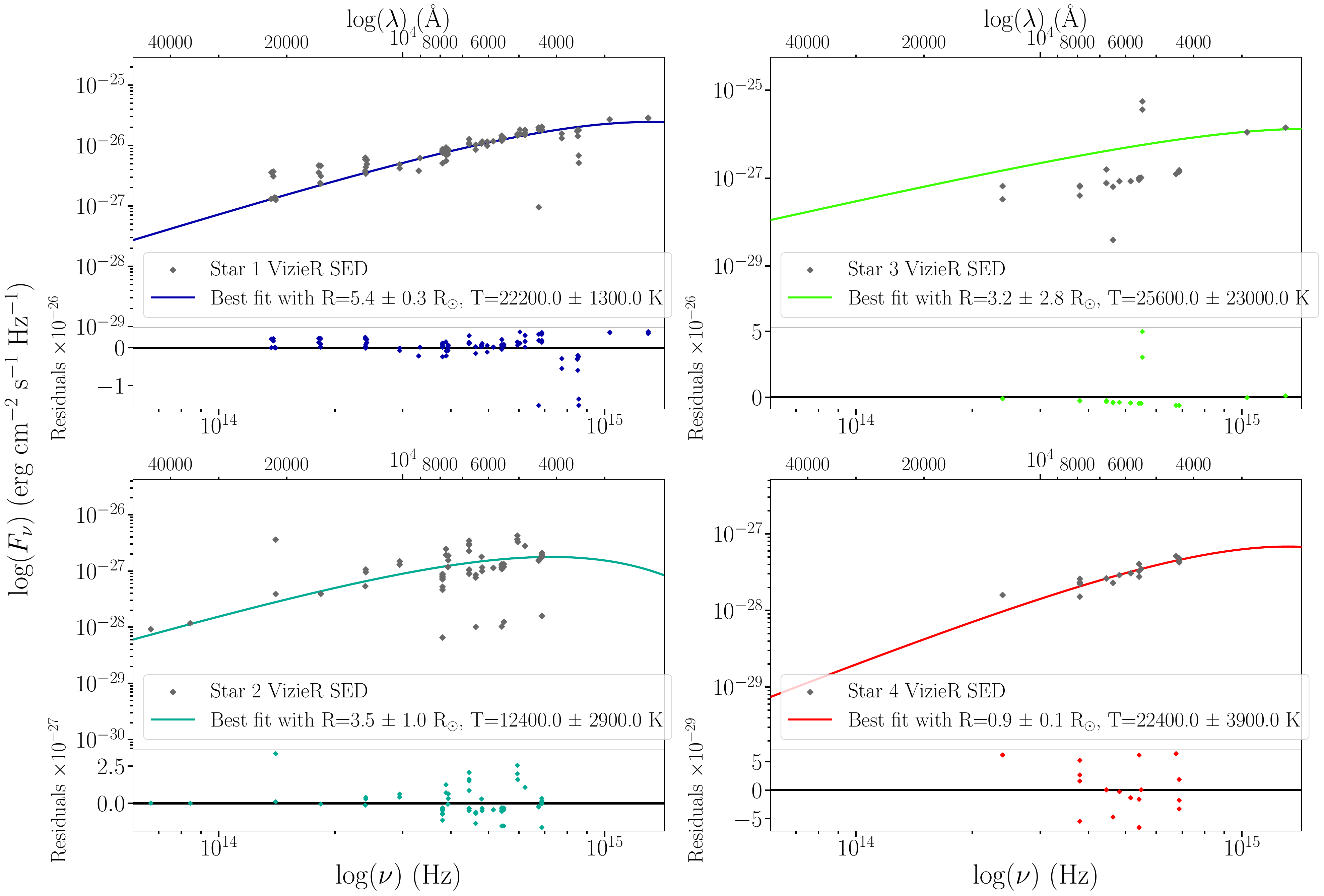}
    \caption{Modified blackbody curve fit for each star near 1SCUBEDX J005606.0-722749. Star 1 is the candidate optical companion to a BeXRB.}
    \label{fig:72fitting}
\end{figure*}

\begin{figure}
    \centering
    \includegraphics[scale=0.35]{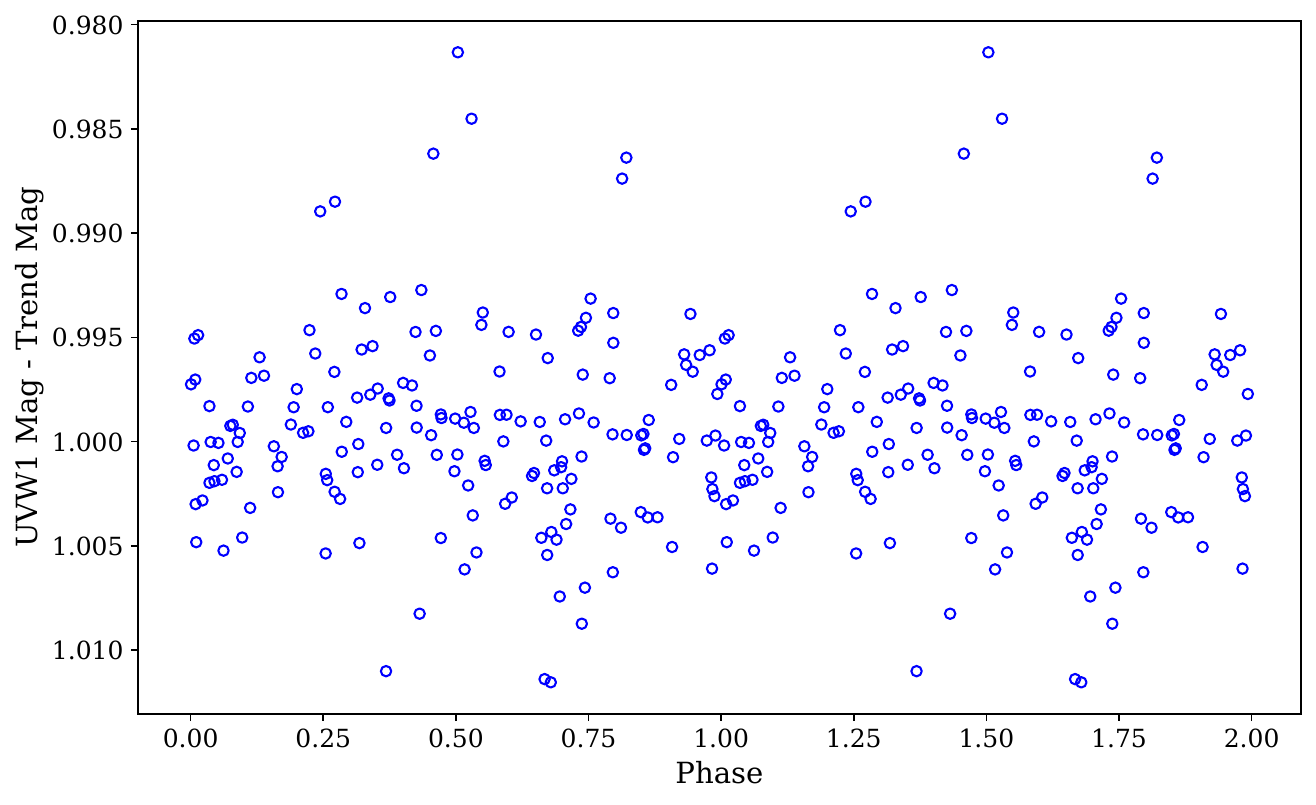}
    \caption{Phase-folded UVOT light curve for the eclipsing binary OGLE SMC-ECL-3357. No evidence can be found of an eclipsing behavior in the UVOT light curve despite a clear eclipse being present in the OGLE light curve published by \citet{2008Schmidtke}.}
    \label{fig:72phasefold}
\end{figure}

1SCUBEDX J005606.0-722749 is located on the edge of the cluster NGC 330, so it resides in a crowded field. The crowded field makes it difficult to neatly separate the photometry of individual sources. After removing duplicate sources, Figure \ref{fig:72clustering} shows four \textit{Gaia}-identified stars that are overlapping with the XRT error region and located in a crowded field at the edge of NGC 330. As seen in Figure \ref{fig:72fitting}, fits to our modified blackbody model are obtained for each of the four stars. Based on the fundamental stellar parameters derived for these stars, Star 1 is identified as the best candidate of these four stars. It produces a best-fitting effective temperature of $22600 \pm 1400$ K and a best-fitting stellar radius of $5.4 \pm 0.3$ R$_{\odot}$. 

Star 1 is close to the XRT position of 1SCUBEDX J005606.0-722749 at a separation of 1.81 arcseconds. The position of this star puts it within a separation of 1 arcsecond of the position of the known OGLE source OGLE-SMC-ECL-3357. This source was first identified as variable by \citet{2008Schmidtke}, but the best analysis of the optical light curve comes from the work of \citet{2021Bodi}, who find that the source is a detached binary with a period of 12.86 days. The source is also reported \citep{2008Schmidtke} to be an eclipsing binary with one eclipse observed by OGLE in each orbit.

The UVOT and XRT light curves for this source are shown in Figure \ref{fig:all lcs}. The source is found to be a persistent, variable UVOT source with a mean magnitude of 14.3 since S-CUBED monitoring began. The weekly cadence of UVOT observations prevents reproduction of the periodicity derived using OGLE data, but there is evidence of strong variability on short timescales in the system as the \textit{uvw1}-band magnitude varies by as much as 0.2 magnitudes from week to week. On longer timescales, the mean magnitude has a slight periodic variation that is similar to the variations observed in Swift J010902.6-723710, but the variation is much less pronounced. There have been 3 isolated XRT detections of this source that suggest a small amount of quiescent X-ray emission is being produced by the NS in the candidate binary. However, a series of low-luminosity Type I outbursts cannot be ruled out as the cause of this emission.

A 12.86 day period is in agreement with the accepted values for a BeXRB. The presence of a narrow eclipse would be a rare occurrence, but there have been examples found of eclipsing BeXRBs. Perhaps the best example of such a system is Swift J010209.6-723710, which was discussed in Section \ref{subsec:Source1772} and described in \citep{2024Gaudin}. Swift J010209.6-723710 shows evidence of a deep eclipse that can be observed at both optical and UV wavelengths every 60.623 days which is interpreted to be the signature of a persistent accretion disk passing in front of the Be star. Other examples include SXP 5.05 \citep{2015Coe}, a system in which X-ray emission from the NS is eclipsed by its companion, and LXP 168.8 \citep{2013Maggi}, which displays similar behavior to Swift J010209.6-723710.

In the case of Swift J010209.6-723710, the eclipse can be observed in both the OGLE and UVOT phase-folded light curves \citep{2024Gaudin}. It is clear that such an eclipse can be observed in the phase-folded OGLE-III light curve of \citet{2008Schmidtke}. A similar phase-folded light curve can be produced from S-CUBED UVOT data by using the ``rspline" method inherent to the W\=otan \citep{2019Hippke} package to de-trend and flatten the data. The resulting phase-folded light curve is shown in Figure \ref{fig:72phasefold}. There is no evidence of an eclipse that is present within this data, preventing the source from being compared to similar sources. It is possible that the accretion disk in this system is much smaller than in other known eclipsing BeXRB systems, leading to shallower eclipses that are below the detection threshold of UVOT. This would be supported by the lack of X-ray detections over the lifespan of the S-CUBED survey. However, it is also possible that the lack of a UVOT detection indicates that this source is indeed not a BeXRB and is in fact some other type of eclipsing binary system. More observations are needed in order to better understand this system.

\subsection{1SCUBEDX J010203.7-715130}\label{subsec:Source131}

\begin{figure}
    \centering
    \includegraphics[scale=0.25]{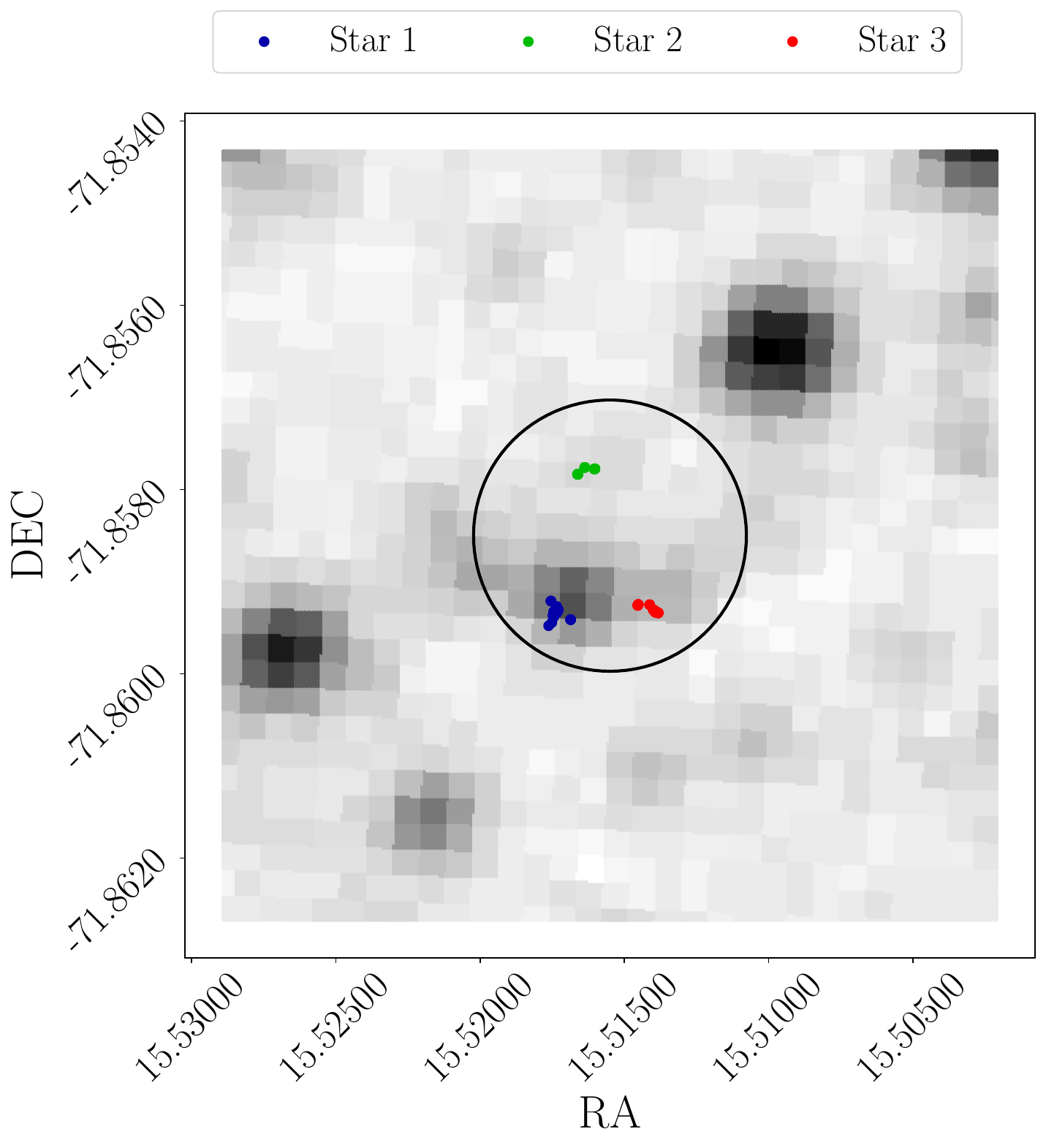}
    \caption{The locations of all photometric data for each star detected by \textit{Gaia} near 1SCUBEDX J010203.7-715130. Photometric positions are plotted over a DSS \citep{2020STScI} reference image for the region. The black circle represents the Swift XRT error region for the source. Five stars are found in close proximity to this source.}
    \label{fig:131clustering}
\end{figure}

\begin{figure}
    \centering
    \includegraphics[scale=0.18]{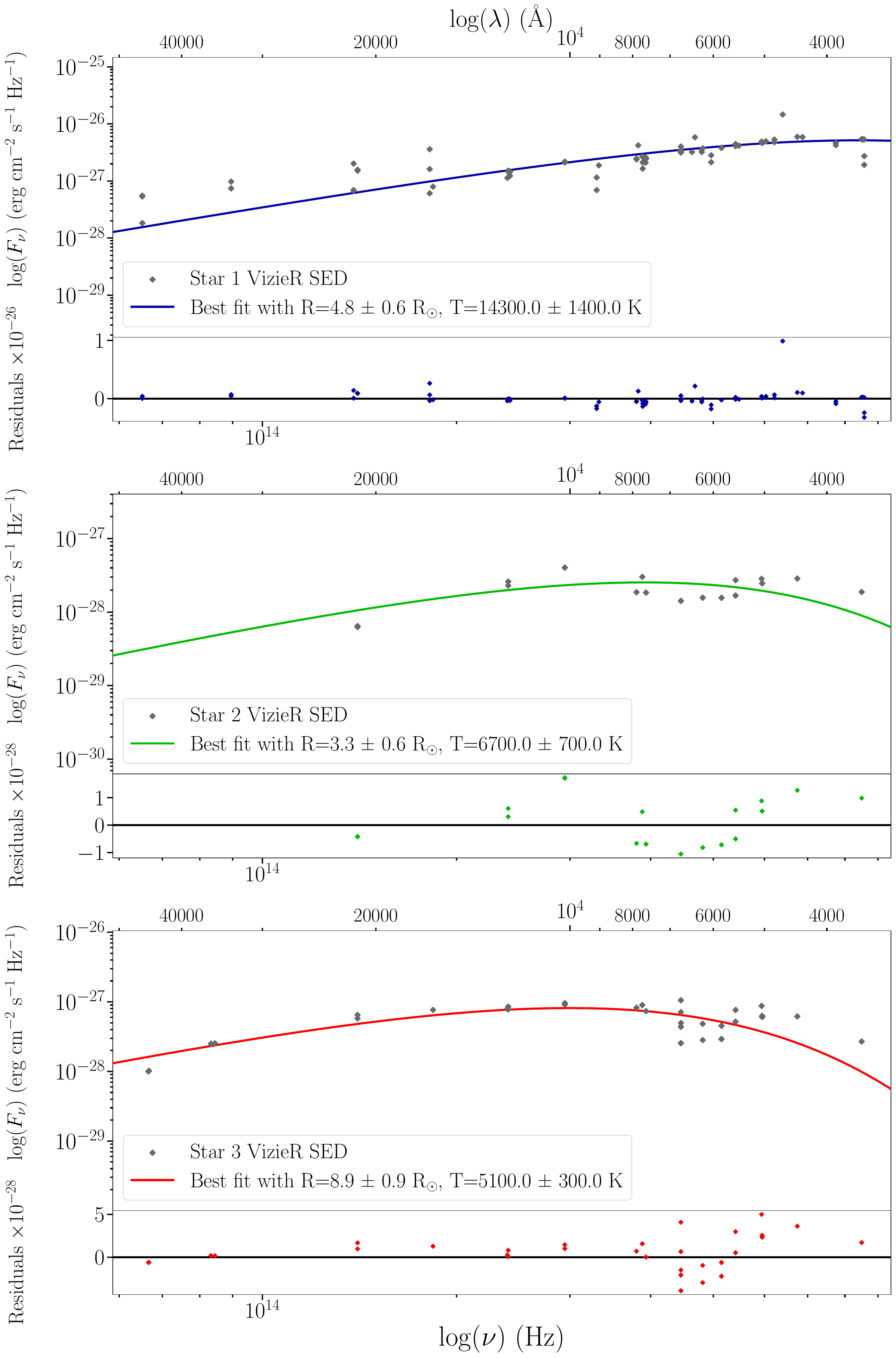}
    \caption{Modified blackbody curve fit for each star near 1SCUBEDX J010203.7-715130. Star 1 is the candidate optical companion to a BeXRB.}
    \label{fig:131fitting}
\end{figure}

The position of 1SCUBEDX J010203.7-715130 places it at the intersection of multiple S-CUBED observation tiles. This is not a problem for XRT as the tiles overlap to provide full coverage, but the source lands outside of UVOT field of view for both tiles. Therefore, it is not possible to produce a UVOT light curve for this source. We must rely exclusively on SED-fitting to evaluate the nature of this candidate source.

Three \textit{Gaia} sources intersect with the XRT error region as shown by Figure \ref{fig:131clustering}. The SED-fitting results for these three stars are shown in Figure \ref{fig:131fitting}. Only one of these stars, Star 1, has the derived fundamental stellar parameters  expected of an OB star in the SMC. This star is located 3.65 arcseconds from the center of the XRT error region and fits to our modified blackbody curve with an effective temperature of $14300 \pm 1400$ K and a stellar radius of $4.8 \pm 0.6$ R$_\odot$, which are within the range that is expected for a relatively cool B-type star. In addition to having stellar parameters corresponding to a B star, the SED shows a similar shape to those found in well-known BeXRB systems (for a discussion, see Section \ref{sec:discussion}) with a strong IR excess clearly visible in the SED. If this is indeed a Be star, then it is likely that the IR excess artificially cools the best-fit for the star and shrinks its radius. Follow-up spectroscopic measurements may be able to better constrain the parameters of the star. 

\subsection{1SCUBEDX J011535.0-731931}\label{subsec:Source151}

\begin{figure}
    \centering
    \includegraphics[scale=0.25]{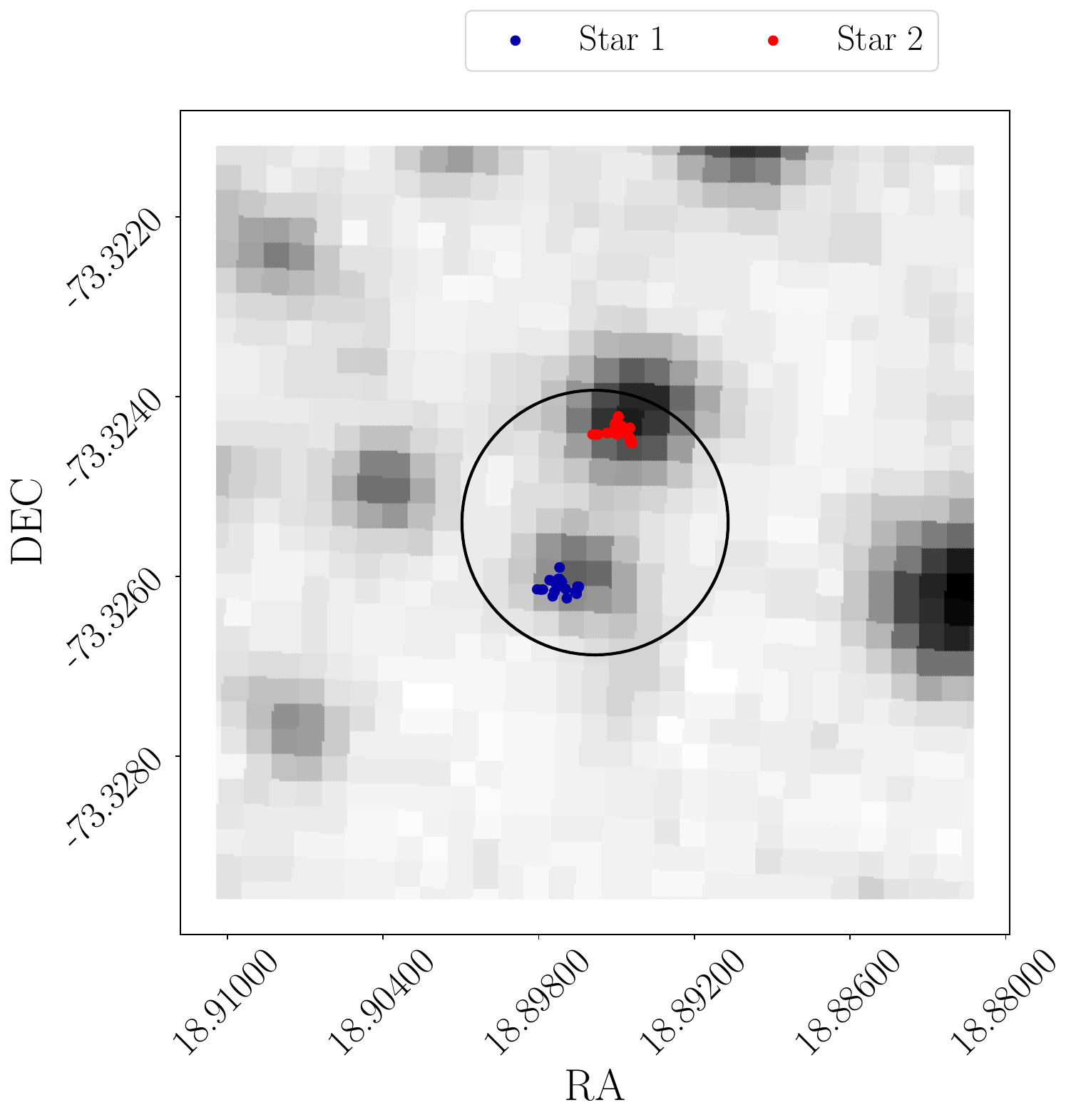}
    \caption{The locations of all photometric data for each star detected by \textit{Gaia} near 1SCUBEDX J011535.0-731931. Photometric positions are plotted over a DSS \citep{2020STScI} reference image for the region. The black circle represents the Swift XRT error region for the source. Only three stars are found in close proximity to this source. }
    \label{fig:151clustering}
\end{figure}

\begin{figure}
    \centering
    \includegraphics[scale=0.17]{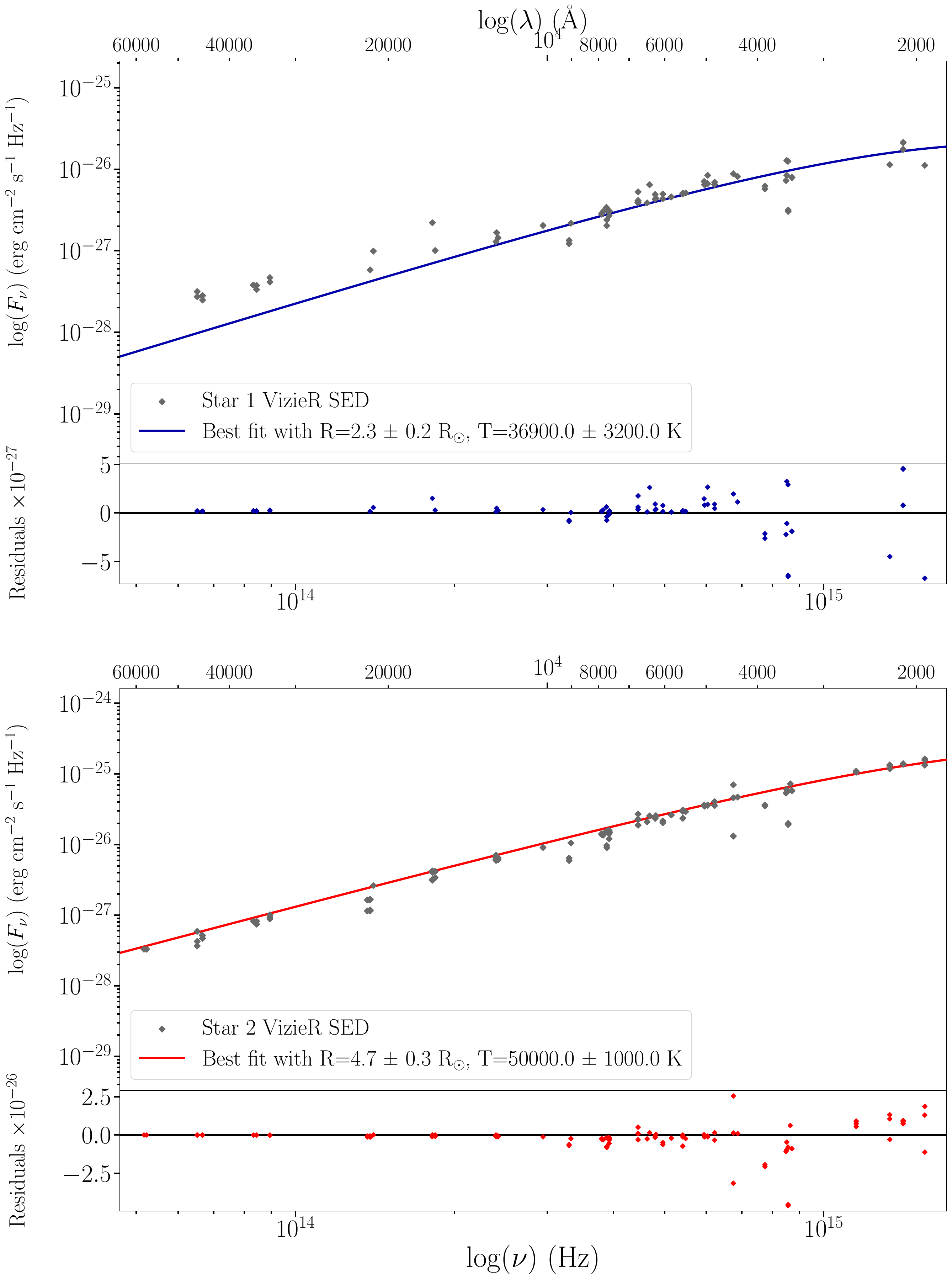}
    \caption{Modified blackbody curve fit for each star near 1SCUBEDX J011535.0-731931. Either star could be the candidate optical companion to a BeXRB.}
    \label{fig:151fitting}
\end{figure}

1SCUBEDX J011535.0-731931 is again located at the edge of multiple S-CUBED XRT tiles, so there is very limited UVOT data available for the source. However, unlike 1SCUBEDX J010203.7-715130, this source is accessible at certain Swift roll angles, leading to sparse UVOT coverage and producing the light curve observed in Figure \ref{fig:all lcs}. 1SCUBEDX J011535.0-731931 has been detected 24 times by S-CUBED UVOT observations, which confirms that the source is a strong, variable UV emitter. However, only 3 of these observations have occurred after MJD 59000. Despite the limited UVOT coverage, this source is a periodic X-ray emitter with 9 evenly-spaced XRT detections over the lifetime of S-CUBED. The most recent detection was a set of 3 observations near MJD 60000 where the source reached its brightest-known X-ray luminosity, but this was still a very faint luminosity of $L_X = 7.0 \times 10^{35}$ erg s$^{-1}$.

SED-fitting for this source is non-conclusive due to multiple candidate optical companions in the field. Figure \ref{fig:151clustering} shows the position of these nearby stars and Figure \ref{fig:151fitting} shows the results of SED-fitting. From fitting, it is evident that two stars are identified by \textit{Gaia}. Star 1 and Star 2 both have an argument to be the optical companion to the candidate BeXRB. Star 1 is located 2.66 arcseconds from the center of the XRT error region. The best-fitting temperature and radius for this star are $36900 \pm 3200$ K and $2.3 \pm 0.2$ R$_\odot$, respectively. While the temperature is within the expected range for Be stars, the small stellar radius would make this the smallest Be star in the SMC by a significant margin. There are also hints of an IR excess present in the SED of this star. Star 2 is located 3.89 arcseconds from the center of the XRT error region, but shows no indication of an IR excess. The parameters derived from SED-fitting for this star indicate a hot OB star, but SED-fitting is inconclusive for the source. The derived radius radius of $4.7 \pm 0.3$ R$_\odot$ is small for a main sequence star given such a high temperature. The derived temperature of $50000 \pm 3400$ K suggests that the peak of the blackbody is not being captured by the fitting code, leading to large uncertainties in the true stellar temperature.  Despite these uncertainties, it is clear that this star has the temperature of an O-type star, so it is likely that the SED-fitting process is underestimating the true stellar radius. Based on the derived stellar parameters and the IR excess of both sources, the parameters of either star would be appropriate for the optical counterpart to a BeXRB. More observations are needed to conclusively determine which star is a better match as the true optical counterpart to the X-ray source and to place better constraints on the stellar parameters.

\subsection{1SCUBEDX J003802.8-734458}\label{subsec:Source251}

\begin{figure}
    \centering
    \includegraphics[scale=0.25]{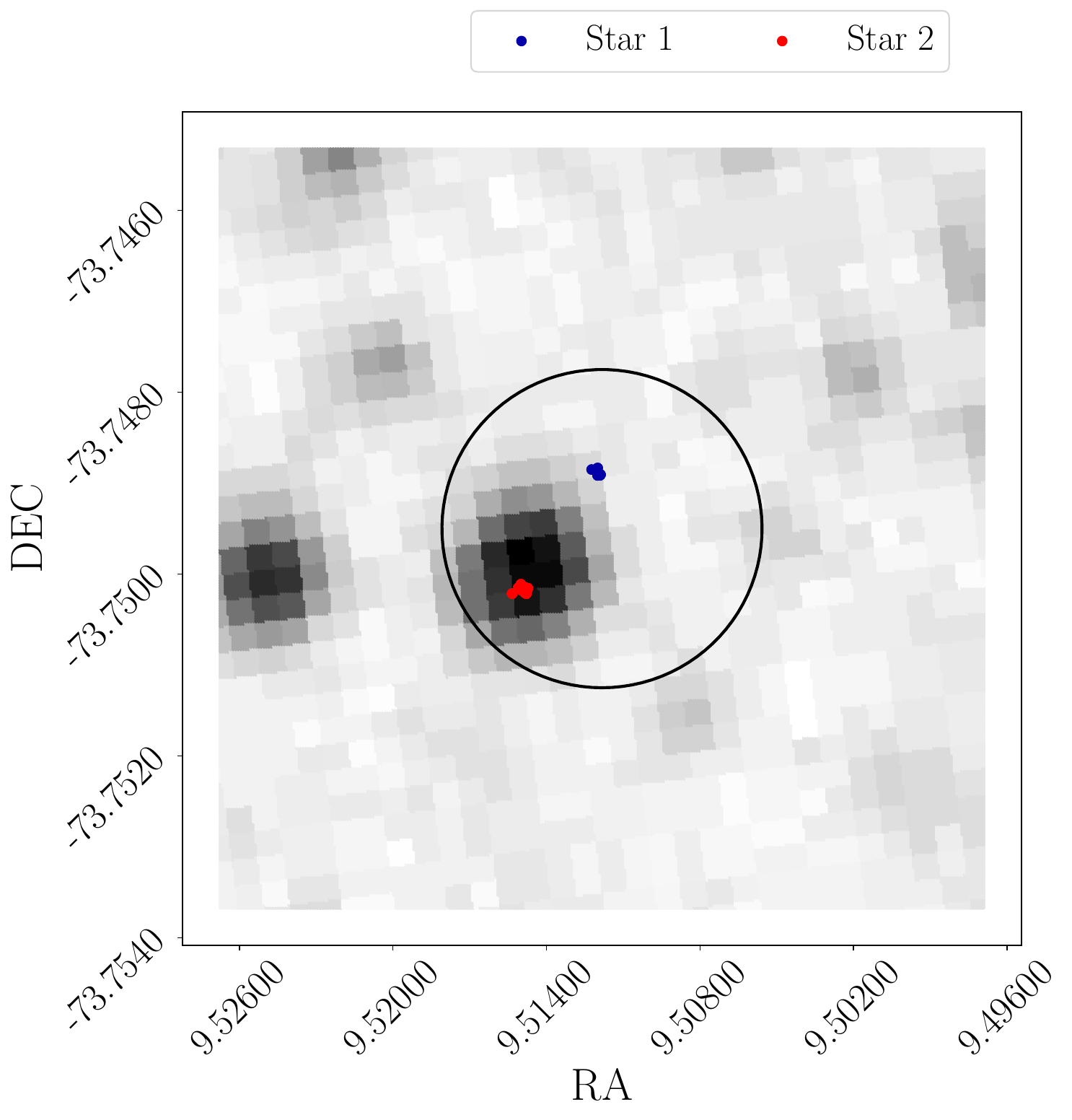}
    \caption{The locations of all photometric data for each star detected by \textit{Gaia} near 1SCUBEDX J003802.8-734458. Photometric positions are plotted over a DSS \citep{2020STScI} reference image for the region. The black circle represents the Swift XRT error region for the source. Two stars are found in close proximity to this source. }
    \label{fig:251clustering}
\end{figure}

\begin{figure}
    \centering
    \includegraphics[scale=0.18]{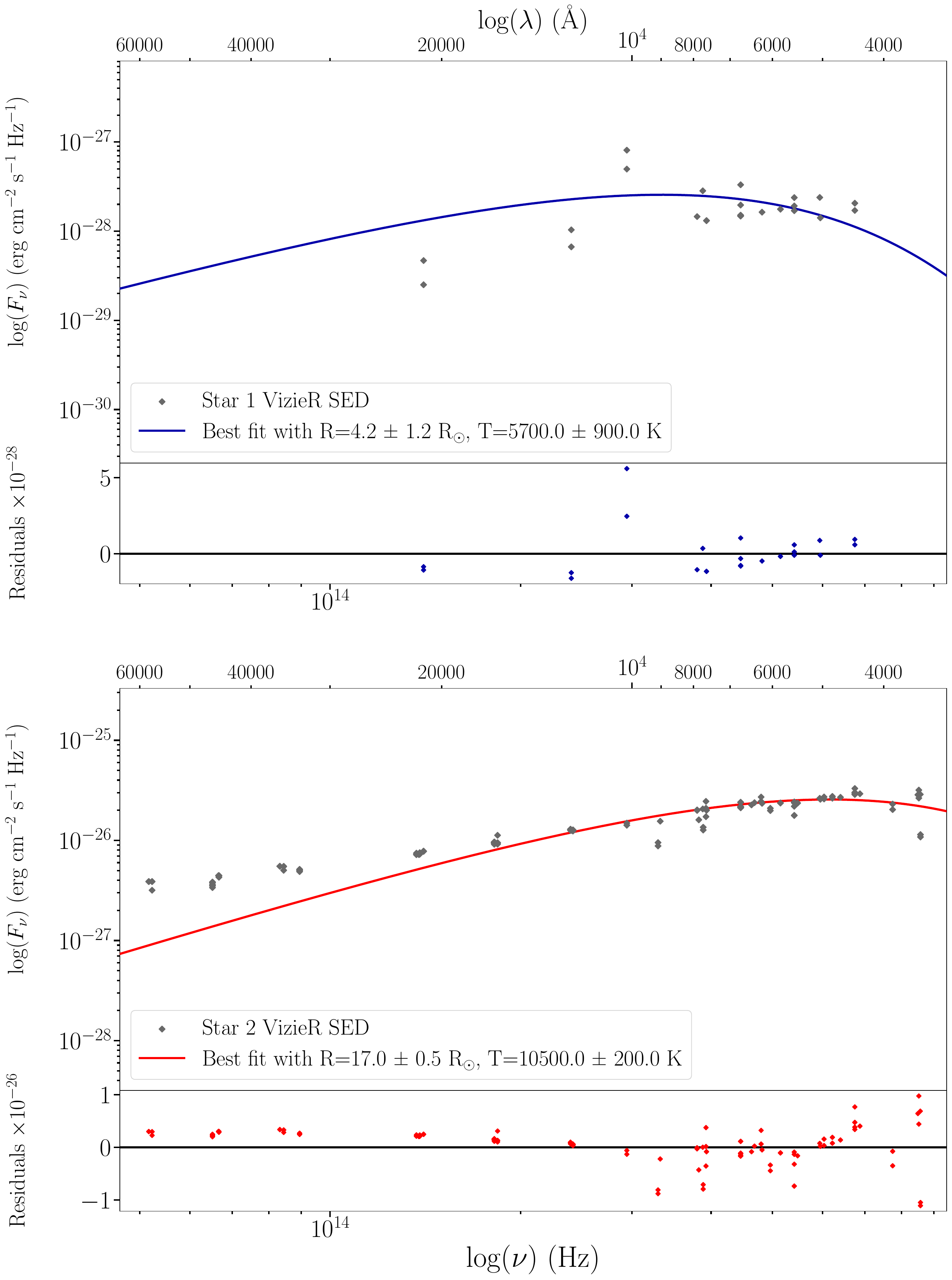}
    \caption{Modified blackbody curve fit for each star near 1SCUBEDX J003802.8-734458. Star 2 is the candidate optical companion to a BeXRB.}
    \label{fig:251fitting}
\end{figure}

1SCUBEDX J003802.8-734458 is the only source besides Swift J010902.6-723710 with a strong-enough X-ray detection to produce a 0.3-10 keV spectrum. Automatic XRT analysis of the S-CUBED spectrum indicates a photon index of $0.8_{-0.6}^{+0.9}$. This fit indicates that the source has a hard X-ray spectrum as is characteristic of a BeXRB. Figure \ref{fig:251clustering} shows that only two \textit{Gaia} sources are found on the periphery of the XRT error region. The fits of both stars to our modified blackbody model are shown in Figure \ref{fig:251fitting}. One of these sources, Star 2, has stellar parameters that correspond to a candidate Be star. Star 2 fits to a stellar blackbody with a temperature of $10500 \pm 200$ K and a radius of $17.0 \pm 0.5$. These parameters indicate that this optical companion would be more evolved than any other optical companion in this study. However, placing this on an HR diagram with known BeXRBs (see Section \ref{sec:discussion}), we can see that these parameters are consistent with an optical OB companion. Additionally, the characteristic IR excess of a Be star appears in the SED for this source.

UVOT observations indicate that there is persistent, strong UV emission that is present over the entire life-span of the S-CUBED survey. The average \textit{uvw1}-band magnitude over all observations is 14.2, but this can vary on short timescales by as much as 0.2 magnitudes. As was the case for 1SCUBEDX J005606.0-722749, there is some evidence of variation on the timescale of $\sim$1.5 years, but this variation is does not shift the mean magnitude by a significant amount as it did in Swift J010902.6-723710 and does not seem to imply that an outburst is imminent. Instead this source seems to be capabale of producing periodic X-ray emission without a major change in the structure of a potential Be star disk. There are 8 XRT detection events since the start of S-CUBED, but these events have produced only weak emission ($<$0.05 counts per second). None of these detection events have been associated with a large rise or drop in UV brightness. Additionally, there have been no XRT detections since MJD 59569, and there has been no change in UV brightness since this event. 

\subsection{1SCUBEDX J005708.8-724202}\label{subsec:Source430}

\begin{figure}
    \centering
    \includegraphics[scale=0.25]{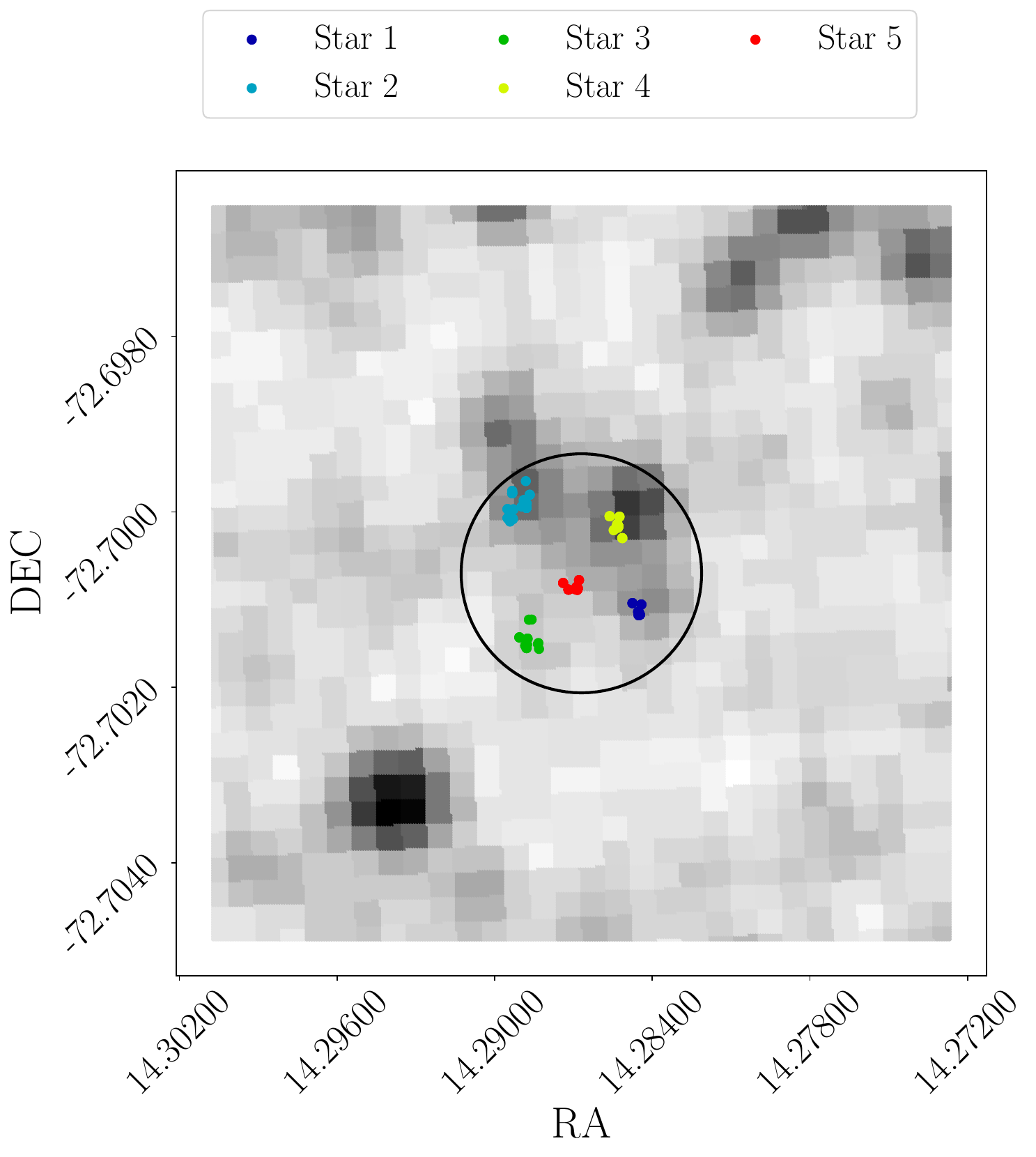}
    \caption{The locations of all photometric data for each star detected by \textit{Gaia} near 1SCUBEDX J005708.8-724202. Photometric positions are plotted over a DSS \citep{2020STScI} reference image for the region. The black circle represents the Swift XRT error region for the source. Five stars are found in close proximity to this source.}
    \label{fig:430clustering}
\end{figure}

\begin{figure*}
    \centering
    \includegraphics[scale=0.23]{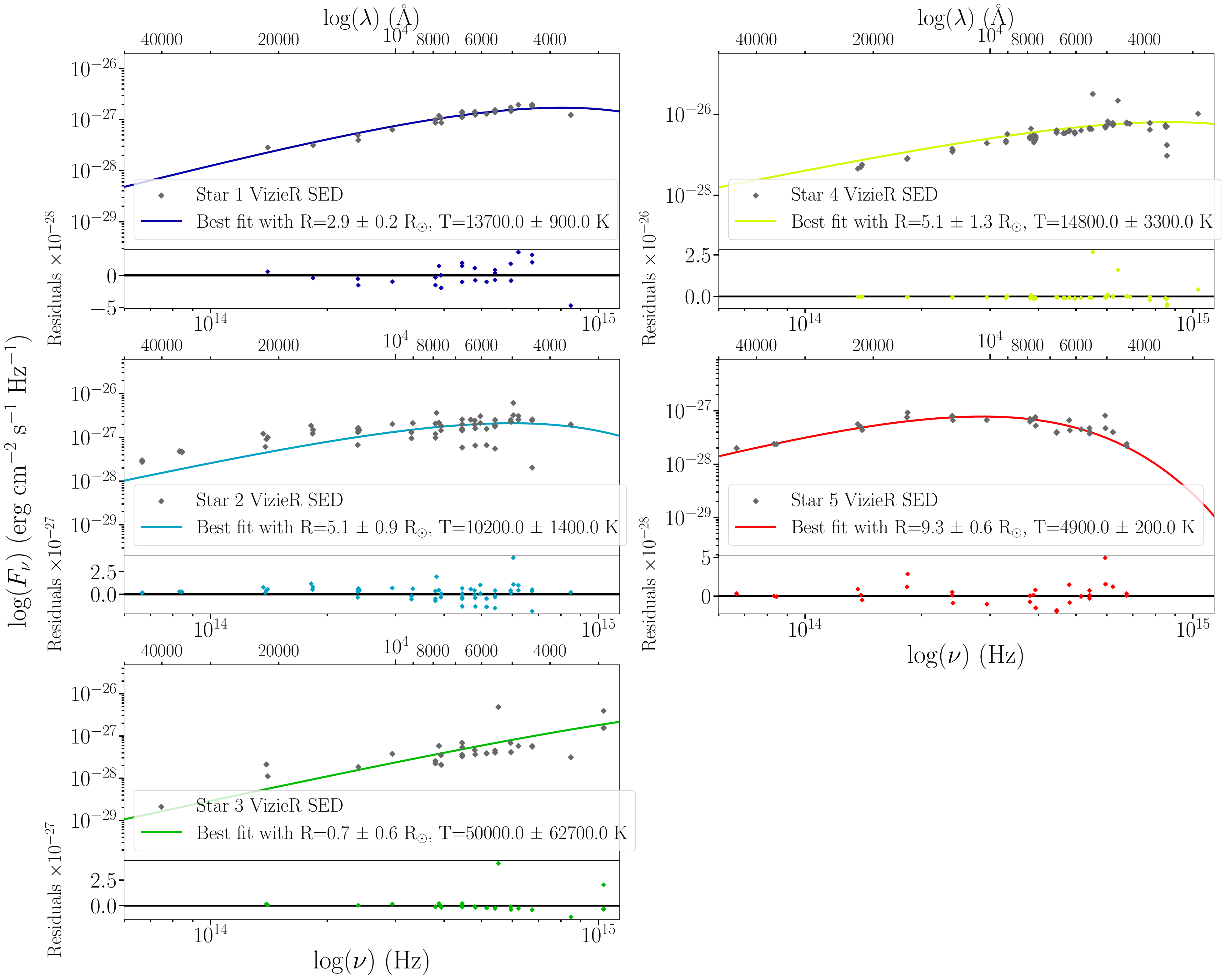}
    \caption{Modified blackbody curve fit for each star near 1SCUBEDX J005708.8-724202. Star 4 is the candidate optical companion to a BeXRB.}
    \label{fig:430fitting}
\end{figure*}

The last source, 1SCUBEDX J005708.8-724202, has perhaps the weakest evidence to be considered a BeXRB. Six stars are found by \textit{Gaia} to insersect the XRT error region, as is seen in Figure \ref{fig:430clustering}. When these stars are fit to our modified blackbody function as shown in Figure \ref{fig:430fitting}, the best candidate Be star of the six is Star 4. The best-fitting temperature for this star is found to be $14800 \pm 3300$ K, and the best-fitting radius is found to be $5.1 \pm 1.3$ R$_\odot$. The evidence for NIR excess in the Star 5 SED is weak, so the disk of the optical companion is likely to be small if this source is indeed a BeXRB. Additionally, this is another source where the radius is smaller than is expected. More observations are needed to constrain the stellar parameters, but the combination of the radius and temperature does place the star in the right region of the HR diagram to be a BeXRB optical companion.

UVOT and XRT light curve data for 1SCUBEDX J005708.8-724202 are presented in Figure \ref{fig:all lcs} and demonstrate that this source is a persistent UV emitter with a mean magnitude of 15.7 in the \textit{uvw1}-band. Unlike the other candidates with UVOT light curves, this source is remarkably stable over long timescales. There is very little evidence of variability on the 500-1000 day timescales that are a characteristic of other candidates discussed in this paper. Despite a lack of variability on long timescales, there is evidence of the same short-timescale variability observed in other candidates as the UV magnitude is capable of varying by as much as 0.2 magnitudes per week. The XRT light curve reveals that this source has only been detected three times with two of those detections occurring in quick succession on October 26, 2016 and December 29, 2016. The peak count rate of this small X-ray flare reached 0.1 counts s$^{-1}$ which could potentially correspond to a series of small, overlooked Type I outbursts. However, this series of X-ray detections was not accompanied by any growth of the UV magnitude as would be expected for a BeXRB outburst. 

\section{Discussion} \label{sec:discussion}

\begin{figure*}
    \centering
    \includegraphics[scale=0.57]{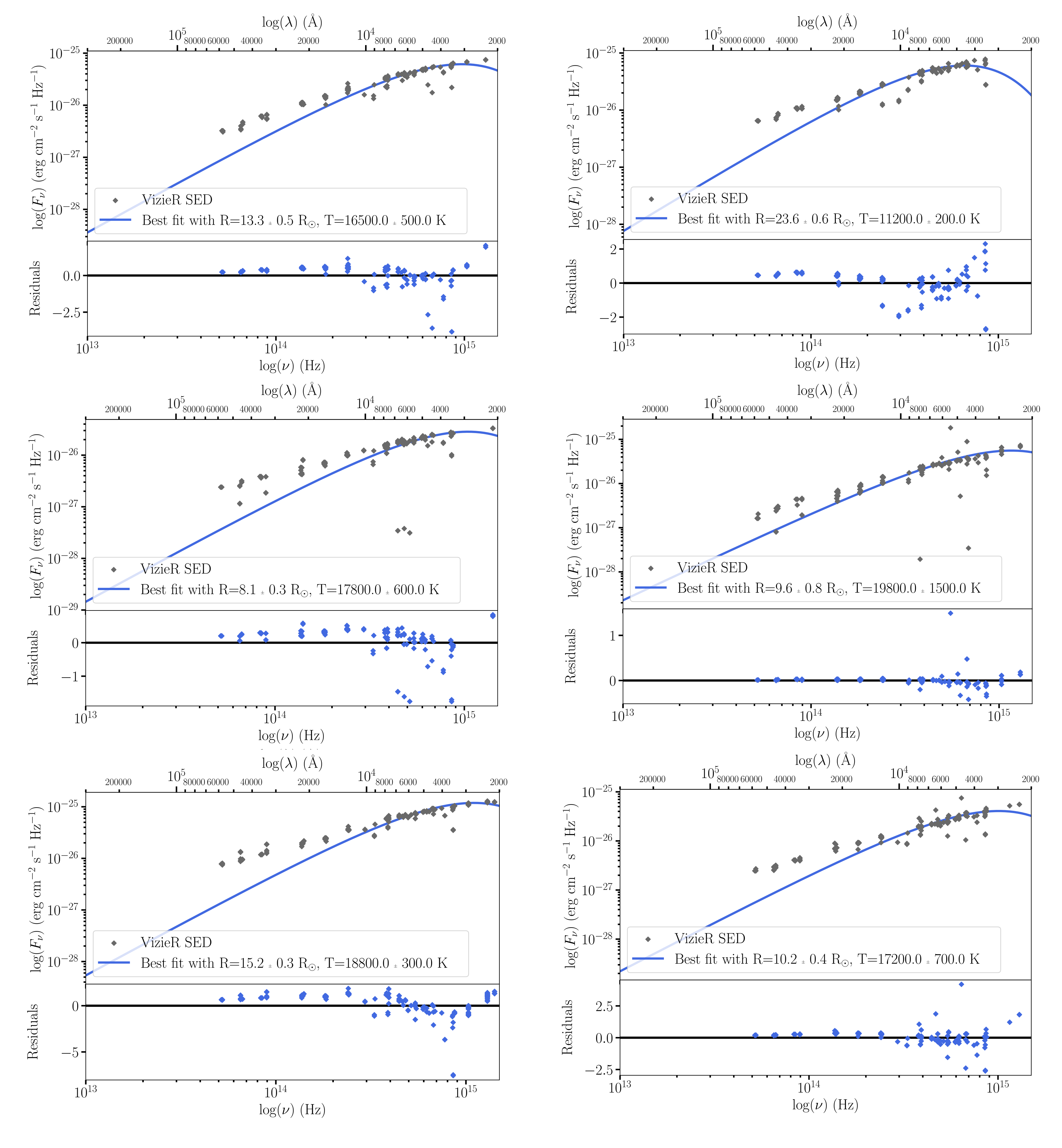}
    \caption{A selected group of SEDs for BeXRBs that have been identified as part of the S-CUBED source catalogue. Each source SED has has been fit using the methods described in Section \ref{subsec:sed} in order to get a temperature and radius for the companion star in each system. The best-fitting stellar blackbody curve and error bars have been plotted for each best-fit line. For all systems with MIR observations available, an emission bump can be observed starting at $\sim 20 \, \mu$m. }
    \label{fig:bexrb fits}
\end{figure*}

\begin{figure}
    \centering
    \includegraphics[scale=0.45]{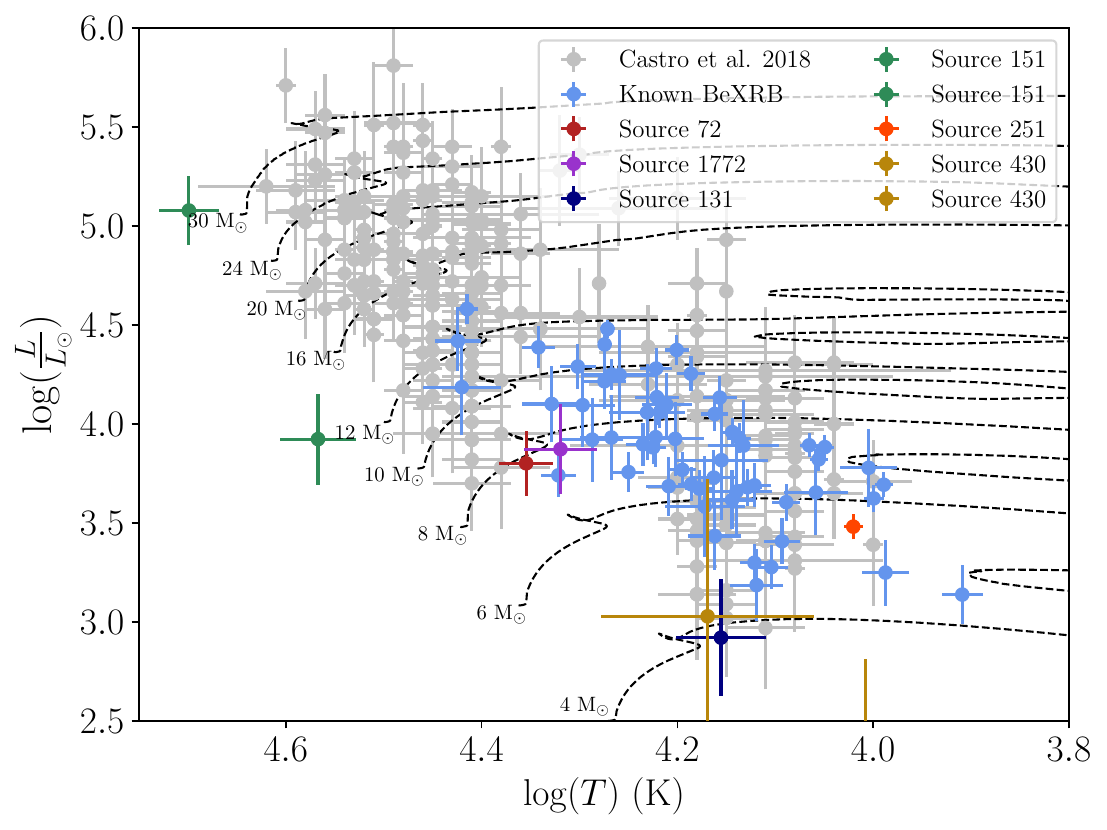}
    \caption{HR Diagram showing stellar parameters derived via SED fitting for the six candidate BeXRBs identified by this study and the population of known BeXRBs within the S-CUBED sample. Also plotted are the stellar parameters for all OB stars within the SMC as derived spectroscopically by \citet{2018Castro} and stellar evolution tracks for high-mass stars in the SMC as computed by MIST \citep{2016Dotter}.}
    \label{fig:hr diagram}
\end{figure}

The methods described in Section \ref{subsec:sed} are shown be capable of detecting B-type stars near candidate systems using SED-fitting. In order to further validate these techniques, the same SED-fitting method can be applied to the population of known BeXRBs that are listed as sources in the S-CUBED catalog. SED-fitting is useful both for the identification of the fundamental stellar parameters of the Be companion in known systems and the features that consistently define the SED of a BeXRB. 

Figure \ref{fig:bexrb fits} shows the fits of our modified blackbody model to several S-CUBED-identified BeXRBs. From these fits, it is evident that the UV and optical data largely follows a stellar blackbody curve. It is also evident that in the IR regime, that dominant emission source is re-processed stellar emission from the Be disk \citep{2011Reig}. The IR SED shows a significant IR excess over what would be expected from a star with no disk. Given how strong this feature can be in some cases, visual evidence of an IR excess present in SED-fitting results is likely one of the best diagnostic tools available for identifying candidate BeXRBs.
% All of these features are predicted by models, however, there are features in the IR that are not predicted by any known model. In many BeXRB systems, a significant excess starting \textbf{at $\sim$10 $\mu$m, }which corresponds to photometric data from the Wide-field Infrared Survey Explorer (WISE). The WISE data shows a significant change in slope of the IR SED curve. Theoretical models predict that the disk SED should follow a power law in the IR regime \citep{2013Rivinius}, but that is not observed here. To the best of our knowledge, this feature has never been identified or discussed in the literature. It may be an instrumental or observational effect. However, if it were shown to be a real feature of SMC BeXRBs, then this significant IR excess would serve as a useful diagnostic for identifying future BeXRBs.

The identification of fundamental stellar parameters allows for an estimate of spectral class and type to be made for each BeXRB optical companion star. When we compare these stars to the larger population of OB stars in the Milky Way, we get a sense of the parameters expected in the population. SED-fitting produces the radius and effective temperature of a given star, which inherently allows for the derivation of the stellar luminosity. Using the derived luminosities and temperatures from SED-fitting, the candidate sources identified by archival analysis and the population of known BeXRBs in the S-CUBED catalog can be placed on an HR diagram. Additionally, an HR diagram of all OB stars in the SMC is created using spectroscopic data from  \citet{2018Castro}. The resulting HR diagram with all three populations is shown in Figure \ref{fig:hr diagram}. Plotted alongside the different populations are theoretical stellar evolution tracks produced by the MESA Isochrones \& Stellar Tracks (MIST; \citealt{2016Dotter}) project. These particular tracks are generated for high-mass stars in a low-metallicity ($[Fe/H] = -1.0$) environment such as is expected for OB stars in the SMC.

The fundamental stellar parameters of the optical companion are not influenced by the presence of a companion. However, plotting our candidate systems alongside these other two populations provides an extra check to ensure that the parameters derived from SED-fitting match the parameters expected for members of known populations. Any candidate with nearby stars that do not fall in the same locus of the HR diagram as known OB and OBe stars is likely not a true BeXRB. All six candidates fall within the locus of the HR diagram containing both SMC OB stars and known BeXRB companions and can be divided into three different groups. Swift J010902.6-723710 and 1SCUBEDX J005606.0-722749 have ``average" stellar parameters for BeXRB companions or OB stars. Neither of these sources is unusual in terms of temperature or radius, and the nearby OB star to 1SCUBEDX J003802.8-734458 is the most similar candidate to the one confirmed BeXRB that has been identified by this study. Three candidates, 1SCUBEDX J010203.7-715130, 1SCUBEDX J003802.8-734458 and 1SCUBEDX J005708.8-724202, have stellar parameters which would indicate a cooler star. The major difference between these three systems is their luminosities. 1SCUBEDX J003802.8-734458 is more luminous than the other two systems, indicating that it may be nearby to an evolved OB-type star. The best-fitting parameters of both stars nearby to 1SCUBEDX J011535.0-731931 represent an outlier on this HR diagram. If either nearby star is an OB star, then they are hot, main sequence stars. However, the uncertainty in the best-fit parameters causes them to be shifted to hotter temperatures than are observed for other systems on this HR diagram.  

\citet{2018Castro} notes that the population of Be stars in their sample appear cooler and more evolved than their spectral type indicates. This phenomenon is attributed by the authors to the presence of emission lines instead of absorption lines in the stellar spectrum of a Be star. This affects the stellar parameters that can be derived from fitting to these spectral lines. Similarly, our method of SED-fitting appears to produce stellar parameters that are indicative of cooler, more evolved companions. In our case, the cause is likely excess IR continuum emission which changes the shape of the SED for a Be star. Additionally, the IR portion of the SED is affected by variability that is produced by the circumstellar disk surrounding the Be star. Both of these characteristics will affect the accuracy of the fit at redder wavelengths but will have a minimal impact on the bluer portion of the SED where the flux will peak for hot, blue stars. More research is needed to determine the best method of obtaining more accurate stellar parameters in systems containing Be stars. 

\section{Conclusion} \label{sec:conclusion} 

In this paper, we present a new method for the identification of candidate BeXRB systems in the SMC. Candidates are first identified using their IR through X-ray properties using searches of archival databases. Once candidates were identified, the fundamental stellar parameters of potential optical counterparts to BeXRBs are identified via SED-fitting using the method of least squares fitting. UV light curves were generated to confirm that the sources were both UV-bright and variable on the timescales expected of BeXRBs.

Using this novel detection method, six candidate BeXRB systems were identified. SED-fitting shows that all but one candidate have fundamental stellar parameters within 3$\sigma$ of the mean values derived for known BeXRB optical companions within the SMC. One of these systems, Swift J010902.6-723710, was first identified by this study and later confirmed to be a BeXRB when it went into outburst in October 2023. Careful monitoring will be needed going forward to search for signs of outburst in the rest of the candidates which would serve as confirmation of binarity.

The methodology outlined in this paper is used to analyze only a small subset of the 1900+ identified S-CUBED sources. Further work may make it possible to characterize even more unknown sources and make significant progress on BeXRB identification in the SMC. 

%% Also note that the akcnowlodgment environment does not support long amounts of text. If you have a lot of people and institutions to acknowledge, do not use this command. Instead, create a new \section{Acknowledgments}.
\section{acknowledgments}

    This work made use of data supplied by the UK Swift Science Data Centre at the University of Leicester. This research has made use of the SIMBAD database, operated at CDS, Strasbourg, France. This research has made use of the VizieR catalogue access tool, CDS, Strasbourg, France (DOI : 10.26093/cds/vizier). The original description of the VizieR service was published in 2000, A\&AS 143, 23. This work made use of Astropy: a community-developed core Python package and an ecosystem of tools and resources for astronomy \citep{astropy:2013, astropy:2018, astropy:2022}. This work has made use of data from the European Space Agency (ESA) mission {\it Gaia} (\url{https://www.cosmos.esa.int/gaia}), processed by the {\it Gaia} Data Processing and Analysis Consortium (DPAC, \url{https://www.cosmos.esa.int/web/gaia/dpac/consortium}). Funding for the DPAC has been provided by national institutions, in particular the institutions participating in the {\it Gaia} Multilateral Agreement. The Digitized Sky Survey was produced at the Space Telescope Science Institute under U.S. Government grant NAG W-2166. The images of these surveys are based on photographic data obtained using the Oschin Schmidt Telescope on Palomar Mountain and the UK Schmidt Telescope. The plates were processed into the present compressed digital form with the permission of these institutions. JAK and TMG acknowledge the support of NASA contract NAS5-00136. We acknowledge the use of public data from the Swift data archive.


\vspace{3mm}
\facilities{Swift (XRT and UVOT), OGLE}


\software{Astropy \citep{astropy:2013, astropy:2018, astropy:2022}}




\bibliography{BeXB}{}

\begin{thebibliography}{}
\expandafter\ifx\csname natexlab\endcsname\relax\def\natexlab#1{#1}\fi
\providecommand{\url}[1]{\href{#1}{#1}}
\providecommand{\dodoi}[1]{doi:~\href{http://doi.org/#1}{\nolinkurl{#1}}}
\providecommand{\doeprint}[1]{\href{http://ascl.net/#1}{\nolinkurl{http://ascl.net/#1}}}
\providecommand{\doarXiv}[1]{\href{https://arxiv.org/abs/#1}{\nolinkurl{https://arxiv.org/abs/#1}}}

\bibitem[{{Antoniou} {et~al.}(2009){Antoniou}, {Zezas}, {Hatzidimitriou}, \& {McDowell}}]{2009Antoniou}
{Antoniou}, V., {Zezas}, A., {Hatzidimitriou}, D., \& {McDowell}, J.~C. 2009, \apj, 697, 1695, \dodoi{10.1088/0004-637X/697/2/1695}

\bibitem[{{Arnaud}(1996)}]{Arnaud96}
{Arnaud}, K.~A. 1996, in Astronomical Society of the Pacific Conference Series, Vol. 101, Astronomical Data Analysis Software and Systems V, ed. G.~H. {Jacoby} \& J.~{Barnes}, 17

\bibitem[{{Astropy Collaboration} {et~al.}(2013){Astropy Collaboration}, {Robitaille}, {Tollerud}, {Greenfield}, {Droettboom}, {Bray}, {Aldcroft}, {Davis}, {Ginsburg}, {Price-Whelan}, {Kerzendorf}, {Conley}, {Crighton}, {Barbary}, {Muna}, {Ferguson}, {Grollier}, {Parikh}, {Nair}, {Unther}, {Deil}, {Woillez}, {Conseil}, {Kramer}, {Turner}, {Singer}, {Fox}, {Weaver}, {Zabalza}, {Edwards}, {Azalee Bostroem}, {Burke}, {Casey}, {Crawford}, {Dencheva}, {Ely}, {Jenness}, {Labrie}, {Lim}, {Pierfederici}, {Pontzen}, {Ptak}, {Refsdal}, {Servillat}, \& {Streicher}}]{astropy:2013}
{Astropy Collaboration}, {Robitaille}, T.~P., {Tollerud}, E.~J., {et~al.} 2013, \aap, 558, A33, \dodoi{10.1051/0004-6361/201322068}

\bibitem[{{Astropy Collaboration} {et~al.}(2018){Astropy Collaboration}, {Price-Whelan}, {Sip{\H{o}}cz}, {G{\"u}nther}, {Lim}, {Crawford}, {Conseil}, {Shupe}, {Craig}, {Dencheva}, {Ginsburg}, {Vand erPlas}, {Bradley}, {P{\'e}rez-Su{\'a}rez}, {de Val-Borro}, {Aldcroft}, {Cruz}, {Robitaille}, {Tollerud}, {Ardelean}, {Babej}, {Bach}, {Bachetti}, {Bakanov}, {Bamford}, {Barentsen}, {Barmby}, {Baumbach}, {Berry}, {Biscani}, {Boquien}, {Bostroem}, {Bouma}, {Brammer}, {Bray}, {Breytenbach}, {Buddelmeijer}, {Burke}, {Calderone}, {Cano Rodr{\'\i}guez}, {Cara}, {Cardoso}, {Cheedella}, {Copin}, {Corrales}, {Crichton}, {D'Avella}, {Deil}, {Depagne}, {Dietrich}, {Donath}, {Droettboom}, {Earl}, {Erben}, {Fabbro}, {Ferreira}, {Finethy}, {Fox}, {Garrison}, {Gibbons}, {Goldstein}, {Gommers}, {Greco}, {Greenfield}, {Groener}, {Grollier}, {Hagen}, {Hirst}, {Homeier}, {Horton}, {Hosseinzadeh}, {Hu}, {Hunkeler}, {Ivezi{\'c}}, {Jain}, {Jenness}, {Kanarek}, {Kendrew}, {Kern}, {Kerzendorf}, {Khvalko}, {King}, {Kirkby}, {Kulkarni},
  {Kumar}, {Lee}, {Lenz}, {Littlefair}, {Ma}, {Macleod}, {Mastropietro}, {McCully}, {Montagnac}, {Morris}, {Mueller}, {Mumford}, {Muna}, {Murphy}, {Nelson}, {Nguyen}, {Ninan}, {N{\"o}the}, {Ogaz}, {Oh}, {Parejko}, {Parley}, {Pascual}, {Patil}, {Patil}, {Plunkett}, {Prochaska}, {Rastogi}, {Reddy Janga}, {Sabater}, {Sakurikar}, {Seifert}, {Sherbert}, {Sherwood-Taylor}, {Shih}, {Sick}, {Silbiger}, {Singanamalla}, {Singer}, {Sladen}, {Sooley}, {Sornarajah}, {Streicher}, {Teuben}, {Thomas}, {Tremblay}, {Turner}, {Terr{\'o}n}, {van Kerkwijk}, {de la Vega}, {Watkins}, {Weaver}, {Whitmore}, {Woillez}, {Zabalza}, \& {Astropy Contributors}}]{astropy:2018}
{Astropy Collaboration}, {Price-Whelan}, A.~M., {Sip{\H{o}}cz}, B.~M., {et~al.} 2018, \aj, 156, 123, \dodoi{10.3847/1538-3881/aabc4f}

\bibitem[{{Astropy Collaboration} {et~al.}(2022){Astropy Collaboration}, {Price-Whelan}, {Lim}, {Earl}, {Starkman}, {Bradley}, {Shupe}, {Patil}, {Corrales}, {Brasseur}, {N{"o}the}, {Donath}, {Tollerud}, {Morris}, {Ginsburg}, {Vaher}, {Weaver}, {Tocknell}, {Jamieson}, {van Kerkwijk}, {Robitaille}, {Merry}, {Bachetti}, {G{"u}nther}, {Aldcroft}, {Alvarado-Montes}, {Archibald}, {B{'o}di}, {Bapat}, {Barentsen}, {Baz{'a}n}, {Biswas}, {Boquien}, {Burke}, {Cara}, {Cara}, {Conroy}, {Conseil}, {Craig}, {Cross}, {Cruz}, {D'Eugenio}, {Dencheva}, {Devillepoix}, {Dietrich}, {Eigenbrot}, {Erben}, {Ferreira}, {Foreman-Mackey}, {Fox}, {Freij}, {Garg}, {Geda}, {Glattly}, {Gondhalekar}, {Gordon}, {Grant}, {Greenfield}, {Groener}, {Guest}, {Gurovich}, {Handberg}, {Hart}, {Hatfield-Dodds}, {Homeier}, {Hosseinzadeh}, {Jenness}, {Jones}, {Joseph}, {Kalmbach}, {Karamehmetoglu}, {Ka{l}uszy{'n}ski}, {Kelley}, {Kern}, {Kerzendorf}, {Koch}, {Kulumani}, {Lee}, {Ly}, {Ma}, {MacBride}, {Maljaars}, {Muna}, {Murphy}, {Norman}, {O'Steen},
  {Oman}, {Pacifici}, {Pascual}, {Pascual-Granado}, {Patil}, {Perren}, {Pickering}, {Rastogi}, {Roulston}, {Ryan}, {Rykoff}, {Sabater}, {Sakurikar}, {Salgado}, {Sanghi}, {Saunders}, {Savchenko}, {Schwardt}, {Seifert-Eckert}, {Shih}, {Jain}, {Shukla}, {Sick}, {Simpson}, {Singanamalla}, {Singer}, {Singhal}, {Sinha}, {Sip{H{o}}cz}, {Spitler}, {Stansby}, {Streicher}, {{{S}}umak}, {Swinbank}, {Taranu}, {Tewary}, {Tremblay}, {Val-Borro}, {Van Kooten}, {Vasovi{'c}}, {Verma}, {de Miranda Cardoso}, {Williams}, {Wilson}, {Winkel}, {Wood-Vasey}, {Xue}, {Yoachim}, {Zhang}, {Zonca}, \& {Astropy Project Contributors}}]{astropy:2022}
{Astropy Collaboration}, {Price-Whelan}, A.~M., {Lim}, P.~L., {et~al.} 2022, \apj, 935, 167, \dodoi{10.3847/1538-4357/ac7c74}

\bibitem[{{Becker} \& {Wolff}(2007)}]{2007Becker}
{Becker}, P.~A., \& {Wolff}, M.~T. 2007, \apj, 654, 435, \dodoi{10.1086/509108}

\bibitem[{{Blackburn} {et~al.}(1999){Blackburn}, {Shaw}, {Payne}, {Hayes}, \& {Heasarc}}]{1999Blackburn}
{Blackburn}, J.~K., {Shaw}, R.~A., {Payne}, H.~E., {Hayes}, J.~J.~E., \& {Heasarc}. 1999, {FTOOLS: A general package of software to manipulate FITS files}, Astrophysics Source Code Library, record ascl:9912.002.
\newblock \doeprint{9912.002}

\bibitem[{{B{\'o}di} \& {Hajdu}(2021)}]{2021Bodi}
{B{\'o}di}, A., \& {Hajdu}, T. 2021, \apjs, 255, 1, \dodoi{10.3847/1538-4365/ac082c}

\bibitem[{{Burrows} {et~al.}(2005){Burrows}, {Hill}, {Nousek}, {Kennea}, {Wells}, {Osborne}, {Abbey}, {Beardmore}, {Mukerjee}, {Short}, {Chincarini}, {Campana}, {Citterio}, {Moretti}, {Pagani}, {Tagliaferri}, {Giommi}, {Capalbi}, {Tamburelli}, {Angelini}, {Cusumano}, {Br{\"a}uninger}, {Burkert}, \& {Hartner}}]{2005Burrows}
{Burrows}, D.~N., {Hill}, J.~E., {Nousek}, J.~A., {et~al.} 2005, \ssr, 120, 165, \dodoi{10.1007/s11214-005-5097-2}

\bibitem[{{Castro} {et~al.}(2018){Castro}, {Oey}, {Fossati}, \& {Langer}}]{2018Castro}
{Castro}, N., {Oey}, M.~S., {Fossati}, L., \& {Langer}, N. 2018, \apj, 868, 57, \dodoi{10.3847/1538-4357/aae6d0}

\bibitem[{{Coe} {et~al.}(2015){Coe}, {Bartlett}, {Bird}, {Haberl}, {Kennea}, {McBride}, {Townsend}, \& {Udalski}}]{2015Coe}
{Coe}, M.~J., {Bartlett}, E.~S., {Bird}, A.~J., {et~al.} 2015, \mnras, 447, 2387, \dodoi{10.1093/mnras/stu2568}

\bibitem[{{Coe} {et~al.}(2021{\natexlab{a}}){Coe}, {Kennea}, {Evans}, {Townsend}, {Udalski}, {Monageng}, \& {Buckley}}]{2021Coe}
{Coe}, M.~J., {Kennea}, J.~A., {Evans}, P.~A., {et~al.} 2021{\natexlab{a}}, \mnras, 504, 1398, \dodoi{10.1093/mnras/stab972}

\bibitem[{{Coe} {et~al.}(2020){Coe}, {Kennea}, {Evans}, \& {Udalski}}]{2020Coe}
{Coe}, M.~J., {Kennea}, J.~A., {Evans}, P.~A., \& {Udalski}, A. 2020, \mnras, 497, L50, \dodoi{10.1093/mnrasl/slaa112}

\bibitem[{{Coe} {et~al.}(2024){Coe}, {Kennea}, {Monageng}, {Townsend}, {Buckley}, {Williams}, {Udalski}, \& {Evans}}]{2024Coe}
{Coe}, M.~J., {Kennea}, J.~A., {Monageng}, I.~M., {et~al.} 2024, \mnras, 528, 7115, \dodoi{10.1093/mnras/stae495}

\bibitem[{{Coe} {et~al.}(2021{\natexlab{b}}){Coe}, {Udalski}, {Kennea}, \& {Evans}}]{2021CoeB}
{Coe}, M.~J., {Udalski}, A., {Kennea}, J.~A., \& {Evans}, P.~A. 2021{\natexlab{b}}, \mnras, 505, 4417, \dodoi{10.1093/mnras/stab1609}

\bibitem[{{Dotter}(2016)}]{2016Dotter}
{Dotter}, A. 2016, \apjs, 222, 8, \dodoi{10.3847/0067-0049/222/1/8}

\bibitem[{{Evans} {et~al.}(2009){Evans}, {Beardmore}, {Page}, {Osborne}, {O'Brien}, {Willingale}, {Starling}, {Burrows}, {Godet}, {Vetere}, {Racusin}, {Goad}, {Wiersema}, {Angelini}, {Capalbi}, {Chincarini}, {Gehrels}, {Kennea}, {Margutti}, {Morris}, {Mountford}, {Pagani}, {Perri}, {Romano}, \& {Tanvir}}]{2009Evans}
{Evans}, P.~A., {Beardmore}, A.~P., {Page}, K.~L., {et~al.} 2009, \mnras, 397, 1177, \dodoi{10.1111/j.1365-2966.2009.14913.x}

\bibitem[{{Evans} {et~al.}(2014){Evans}, {Osborne}, {Beardmore}, {Page}, {Willingale}, {Mountford}, {Pagani}, {Burrows}, {Kennea}, {Perri}, {Tagliaferri}, \& {Gehrels}}]{2014Evans}
{Evans}, P.~A., {Osborne}, J.~P., {Beardmore}, A.~P., {et~al.} 2014, \apjs, 210, 8, \dodoi{10.1088/0067-0049/210/1/8}

\bibitem[{{Evans} {et~al.}(2020){Evans}, {Page}, {Osborne}, {Beardmore}, {Willingale}, {Burrows}, {Kennea}, {Perri}, {Capalbi}, {Tagliaferri}, \& {Cenko}}]{2020Evans}
{Evans}, P.~A., {Page}, K.~L., {Osborne}, J.~P., {et~al.} 2020, \apjs, 247, 54, \dodoi{10.3847/1538-4365/ab7db9}

\bibitem[{{Fitzpatrick}(1999)}]{1999Fitzpatrick}
{Fitzpatrick}, E.~L. 1999, \pasp, 111, 63, \dodoi{10.1086/316293}

\bibitem[{{Gaia Collaboration} {et~al.}(2016){Gaia Collaboration}, {Prusti}, {de Bruijne}, {Brown}, {Vallenari}, {Babusiaux}, {Bailer-Jones}, {Bastian}, {Biermann}, {Evans}, {Eyer}, {Jansen}, {Jordi}, {Klioner}, {Lammers}, {Lindegren}, {Luri}, {Mignard}, {Milligan}, {Panem}, {Poinsignon}, {Pourbaix}, {Randich}, {Sarri}, {Sartoretti}, {Siddiqui}, {Soubiran}, {Valette}, {van Leeuwen}, {Walton}, {Aerts}, {Arenou}, {Cropper}, {Drimmel}, {H{\o}g}, {Katz}, {Lattanzi}, {O'Mullane}, {Grebel}, {Holland}, {Huc}, {Passot}, {Bramante}, {Cacciari}, {Casta{\~n}eda}, {Chaoul}, {Cheek}, {De Angeli}, {Fabricius}, {Guerra}, {Hern{\'a}ndez}, {Jean-Antoine-Piccolo}, {Masana}, {Messineo}, {Mowlavi}, {Nienartowicz}, {Ord{\'o}{\~n}ez-Blanco}, {Panuzzo}, {Portell}, {Richards}, {Riello}, {Seabroke}, {Tanga}, {Th{\'e}venin}, {Torra}, {Els}, {Gracia-Abril}, {Comoretto}, {Garcia-Reinaldos}, {Lock}, {Mercier}, {Altmann}, {Andrae}, {Astraatmadja}, {Bellas-Velidis}, {Benson}, {Berthier}, {Blomme}, {Busso}, {Carry}, {Cellino}, {Clementini},
  {Cowell}, {Creevey}, {Cuypers}, {Davidson}, {De Ridder}, {de Torres}, {Delchambre}, {Dell'Oro}, {Ducourant}, {Fr{\'e}mat}, {Garc{\'\i}a-Torres}, {Gosset}, {Halbwachs}, {Hambly}, {Harrison}, {Hauser}, {Hestroffer}, {Hodgkin}, {Huckle}, {Hutton}, {Jasniewicz}, {Jordan}, {Kontizas}, {Korn}, {Lanzafame}, {Manteiga}, {Moitinho}, {Muinonen}, {Osinde}, {Pancino}, {Pauwels}, {Petit}, {Recio-Blanco}, {Robin}, {Sarro}, {Siopis}, {Smith}, {Smith}, {Sozzetti}, {Thuillot}, {van Reeven}, {Viala}, {Abbas}, {Abreu Aramburu}, {Accart}, {Aguado}, {Allan}, {Allasia}, {Altavilla}, {{\'A}lvarez}, {Alves}, {Anderson}, {Andrei}, {Anglada Varela}, {Antiche}, {Antoja}, {Ant{\'o}n}, {Arcay}, {Atzei}, {Ayache}, {Bach}, {Baker}, {Balaguer-N{\'u}{\~n}ez}, {Barache}, {Barata}, {Barbier}, {Barblan}, {Baroni}, {Barrado y Navascu{\'e}s}, {Barros}, {Barstow}, {Becciani}, {Bellazzini}, {Bellei}, {Bello Garc{\'\i}a}, {Belokurov}, {Bendjoya}, {Berihuete}, {Bianchi}, {Bienaym{\'e}}, {Billebaud}, {Blagorodnova}, {Blanco-Cuaresma}, {Boch},
  {Bombrun}, {Borrachero}, {Bouquillon}, {Bourda}, {Bouy}, {Bragaglia}, {Breddels}, {Brouillet}, {Br{\"u}semeister}, {Bucciarelli}, {Budnik}, {Burgess}, {Burgon}, {Burlacu}, {Busonero}, {Buzzi}, {Caffau}, {Cambras}, {Campbell}, {Cancelliere}, {Cantat-Gaudin}, {Carlucci}, {Carrasco}, {Castellani}, {Charlot}, {Charnas}, {Charvet}, {Chassat}, {Chiavassa}, {Clotet}, {Cocozza}, {Collins}, {Collins}, {Costigan}, {Crifo}, {Cross}, {Crosta}, {Crowley}, {Dafonte}, {Damerdji}, {Dapergolas}, {David}, {David}, {De Cat}, {de Felice}, {de Laverny}, {De Luise}, {De March}, {de Martino}, {de Souza}, {Debosscher}, {del Pozo}, {Delbo}, {Delgado}, {Delgado}, {di Marco}, {Di Matteo}, {Diakite}, {Distefano}, {Dolding}, {Dos Anjos}, {Drazinos}, {Dur{\'a}n}, {Dzigan}, {Ecale}, {Edvardsson}, {Enke}, {Erdmann}, {Escolar}, {Espina}, {Evans}, {Eynard Bontemps}, {Fabre}, {Fabrizio}, {Faigler}, {Falc{\~a}o}, {Farr{\`a}s Casas}, {Faye}, {Federici}, {Fedorets}, {Fern{\'a}ndez-Hern{\'a}ndez}, {Fernique}, {Fienga}, {Figueras}, {Filippi},
  {Findeisen}, {Fonti}, {Fouesneau}, {Fraile}, {Fraser}, {Fuchs}, {Furnell}, {Gai}, {Galleti}, {Galluccio}, {Garabato}, {Garc{\'\i}a-Sedano}, {Gar{\'e}}, {Garofalo}, {Garralda}, {Gavras}, {Gerssen}, {Geyer}, {Gilmore}, {Girona}, {Giuffrida}, {Gomes}, {Gonz{\'a}lez-Marcos}, {Gonz{\'a}lez-N{\'u}{\~n}ez}, {Gonz{\'a}lez-Vidal}, {Granvik}, {Guerrier}, {Guillout}, {Guiraud}, {G{\'u}rpide}, {Guti{\'e}rrez-S{\'a}nchez}, {Guy}, {Haigron}, {Hatzidimitriou}, {Haywood}, {Heiter}, {Helmi}, {Hobbs}, {Hofmann}, {Holl}, {Holland}, {Hunt}, {Hypki}, {Icardi}, {Irwin}, {Jevardat de Fombelle}, {Jofr{\'e}}, {Jonker}, {Jorissen}, {Julbe}, {Karampelas}, {Kochoska}, {Kohley}, {Kolenberg}, {Kontizas}, {Koposov}, {Kordopatis}, {Koubsky}, {Kowalczyk}, {Krone-Martins}, {Kudryashova}, {Kull}, {Bachchan}, {Lacoste-Seris}, {Lanza}, {Lavigne}, {Le Poncin-Lafitte}, {Lebreton}, {Lebzelter}, {Leccia}, {Leclerc}, {Lecoeur-Taibi}, {Lemaitre}, {Lenhardt}, {Leroux}, {Liao}, {Licata}, {Lindstr{\o}m}, {Lister}, {Livanou}, {Lobel}, {L{\"o}ffler},
  {L{\'o}pez}, {Lopez-Lozano}, {Lorenz}, {Loureiro}, {MacDonald}, {Magalh{\~a}es Fernandes}, {Managau}, {Mann}, {Mantelet}, {Marchal}, {Marchant}, {Marconi}, {Marie}, {Marinoni}, {Marrese}, {Marschalk{\'o}}, {Marshall}, {Mart{\'\i}n-Fleitas}, {Martino}, {Mary}, {Matijevi{\v{c}}}, {Mazeh}, {McMillan}, {Messina}, {Mestre}, {Michalik}, {Millar}, {Miranda}, {Molina}, {Molinaro}, {Molinaro}, {Moln{\'a}r}, {Moniez}, {Montegriffo}, {Monteiro}, {Mor}, {Mora}, {Morbidelli}, {Morel}, {Morgenthaler}, {Morley}, {Morris}, {Mulone}, {Muraveva}, {Musella}, {Narbonne}, {Nelemans}, {Nicastro}, {Noval}, {Ord{\'e}novic}, {Ordieres-Mer{\'e}}, {Osborne}, {Pagani}, {Pagano}, {Pailler}, {Palacin}, {Palaversa}, {Parsons}, {Paulsen}, {Pecoraro}, {Pedrosa}, {Pentik{\"a}inen}, {Pereira}, {Pichon}, {Piersimoni}, {Pineau}, {Plachy}, {Plum}, {Poujoulet}, {Pr{\v{s}}a}, {Pulone}, {Ragaini}, {Rago}, {Rambaux}, {Ramos-Lerate}, {Ranalli}, {Rauw}, {Read}, {Regibo}, {Renk}, {Reyl{\'e}}, {Ribeiro}, {Rimoldini}, {Ripepi}, {Riva}, {Rixon},
  {Roelens}, {Romero-G{\'o}mez}, {Rowell}, {Royer}, {Rudolph}, {Ruiz-Dern}, {Sadowski}, {Sagrist{\`a} Sell{\'e}s}, {Sahlmann}, {Salgado}, {Salguero}, {Sarasso}, {Savietto}, {Schnorhk}, {Schultheis}, {Sciacca}, {Segol}, {Segovia}, {Segransan}, {Serpell}, {Shih}, {Smareglia}, {Smart}, {Smith}, {Solano}, {Solitro}, {Sordo}, {Soria Nieto}, {Souchay}, {Spagna}, {Spoto}, {Stampa}, {Steele}, {Steidelm{\"u}ller}, {Stephenson}, {Stoev}, {Suess}, {S{\"u}veges}, {Surdej}, {Szabados}, {Szegedi-Elek}, {Tapiador}, {Taris}, {Tauran}, {Taylor}, {Teixeira}, {Terrett}, {Tingley}, {Trager}, {Turon}, {Ulla}, {Utrilla}, {Valentini}, {van Elteren}, {Van Hemelryck}, {van Leeuwen}, {Varadi}, {Vecchiato}, {Veljanoski}, {Via}, {Vicente}, {Vogt}, {Voss}, {Votruba}, {Voutsinas}, {Walmsley}, {Weiler}, {Weingrill}, {Werner}, {Wevers}, {Whitehead}, {Wyrzykowski}, {Yoldas}, {{\v{Z}}erjal}, {Zucker}, {Zurbach}, {Zwitter}, {Alecu}, {Allen}, {Allende Prieto}, {Amorim}, {Anglada-Escud{\'e}}, {Arsenijevic}, {Azaz}, {Balm}, {Beck}, {Bernstein},
  {Bigot}, {Bijaoui}, {Blasco}, {Bonfigli}, {Bono}, {Boudreault}, {Bressan}, {Brown}, {Brunet}, {Bunclark}, {Buonanno}, {Butkevich}, {Carret}, {Carrion}, {Chemin}, {Ch{\'e}reau}, {Corcione}, {Darmigny}, {de Boer}, {de Teodoro}, {de Zeeuw}, {Delle Luche}, {Domingues}, {Dubath}, {Fodor}, {Fr{\'e}zouls}, {Fries}, {Fustes}, {Fyfe}, {Gallardo}, {Gallegos}, {Gardiol}, {Gebran}, {Gomboc}, {G{\'o}mez}, {Grux}, {Gueguen}, {Heyrovsky}, {Hoar}, {Iannicola}, {Isasi Parache}, {Janotto}, {Joliet}, {Jonckheere}, {Keil}, {Kim}, {Klagyivik}, {Klar}, {Knude}, {Kochukhov}, {Kolka}, {Kos}, {Kutka}, {Lainey}, {LeBouquin}, {Liu}, {Loreggia}, {Makarov}, {Marseille}, {Martayan}, {Martinez-Rubi}, {Massart}, {Meynadier}, {Mignot}, {Munari}, {Nguyen}, {Nordlander}, {Ocvirk}, {O'Flaherty}, {Olias Sanz}, {Ortiz}, {Osorio}, {Oszkiewicz}, {Ouzounis}, {Palmer}, {Park}, {Pasquato}, {Peltzer}, {Peralta}, {P{\'e}turaud}, {Pieniluoma}, {Pigozzi}, {Poels}, {Prat}, {Prod'homme}, {Raison}, {Rebordao}, {Risquez}, {Rocca-Volmerange}, {Rosen},
  {Ruiz-Fuertes}, {Russo}, {Sembay}, {Serraller Vizcaino}, {Short}, {Siebert}, {Silva}, {Sinachopoulos}, {Slezak}, {Soffel}, {Sosnowska}, {Strai{\v{z}}ys}, {ter Linden}, {Terrell}, {Theil}, {Tiede}, {Troisi}, {Tsalmantza}, {Tur}, {Vaccari}, {Vachier}, {Valles}, {Van Hamme}, {Veltz}, {Virtanen}, {Wallut}, {Wichmann}, {Wilkinson}, {Ziaeepour}, \& {Zschocke}}]{2016GaiaCollaboration}
{Gaia Collaboration}, {Prusti}, T., {de Bruijne}, J.~H.~J., {et~al.} 2016, \aap, 595, A1, \dodoi{10.1051/0004-6361/201629272}

\bibitem[{{Gaudin} {et~al.}(2024){Gaudin}, {Kennea}, {Coe}, {Monageng}, {Udalski}, {Townsend}, {Buckley}, \& {Evans}}]{2024Gaudin}
{Gaudin}, T.~M., {Kennea}, J.~A., {Coe}, M.~J., {et~al.} 2024, \apjl, 965, L10, \dodoi{10.3847/2041-8213/ad354a}

\bibitem[{{G{\'o}rski} {et~al.}(2020){G{\'o}rski}, {Zgirski}, {Pietrzy{\'n}ski}, {Gieren}, {Wielg{\'o}rski}, {Graczyk}, {Kudritzki}, {Pilecki}, {Narloch}, {Karczmarek}, {Suchomska}, \& {Taormina}}]{2020Gorski}
{G{\'o}rski}, M., {Zgirski}, B., {Pietrzy{\'n}ski}, G., {et~al.} 2020, \apj, 889, 179, \dodoi{10.3847/1538-4357/ab65ed}

\bibitem[{{Graczyk} {et~al.}(2020){Graczyk}, {Pietrzy{\'n}ski}, {Thompson}, {Gieren}, {Zgirski}, {Villanova}, {G{\'o}rski}, {Wielg{\'o}rski}, {Karczmarek}, {Narloch}, {Pilecki}, {Taormina}, {Smolec}, {Suchomska}, {Gallenne}, {Nardetto}, {Storm}, {Kudritzki}, {Ka{\l}uszy{\'n}ski}, \& {Pych}}]{2020Graczyk}
{Graczyk}, D., {Pietrzy{\'n}ski}, G., {Thompson}, I.~B., {et~al.} 2020, \apj, 904, 13, \dodoi{10.3847/1538-4357/abbb2b}

\bibitem[{{Haberl} {et~al.}(2000){Haberl}, {Filipovi{\'c}}, {Pietsch}, \& {Kahabka}}]{2000Haberl}
{Haberl}, F., {Filipovi{\'c}}, M.~D., {Pietsch}, W., \& {Kahabka}, P. 2000, \aaps, 142, 41, \dodoi{10.1051/aas:2000136}

\bibitem[{{Harris} \& {Zaritsky}(2004)}]{2004Harris}
{Harris}, J., \& {Zaritsky}, D. 2004, \aj, 127, 1531, \dodoi{10.1086/381953}

\bibitem[{{Hippke} {et~al.}(2019){Hippke}, {David}, {Mulders}, \& {Heller}}]{2019Hippke}
{Hippke}, M., {David}, T.~J., {Mulders}, G.~D., \& {Heller}, R. 2019, \aj, 158, 143, \dodoi{10.3847/1538-3881/ab3984}

\bibitem[{{Ivanov} {et~al.}(2024){Ivanov}, {Cioni}, {Dennefeld}, {de Grijs}, {Craig}, {van Loon}, {Pennock}, {Maitra}, \& {Haberl}}]{2024Ivanov}
{Ivanov}, V.~D., {Cioni}, M.-R.~L., {Dennefeld}, M., {et~al.} 2024, \aap, 687, A16, \dodoi{10.1051/0004-6361/202346504}

\bibitem[{{Kahabka} \& {Pietsch}(1996)}]{1996Kahabka}
{Kahabka}, P., \& {Pietsch}, W. 1996, \aap, 312, 919, \dodoi{10.48550/arXiv.astro-ph/9706075}

\bibitem[{{Kennea} {et~al.}(2020){Kennea}, {Coe}, {Evans}, {Monageng}, {Townsend}, {Siegel}, {Udalski}, \& {Buckley}}]{2020Kennea}
{Kennea}, J.~A., {Coe}, M.~J., {Evans}, P.~A., {et~al.} 2020, \mnras, 499, L41, \dodoi{10.1093/mnrasl/slaa154}

\bibitem[{{Kennea} {et~al.}(2021){Kennea}, {Coe}, {Evans}, {Townsend}, {Campbell}, \& {Udalski}}]{2021Kennea}
---. 2021, \mnras, 508, 781, \dodoi{10.1093/mnras/stab2632}

\bibitem[{{Kennea} {et~al.}(2018){Kennea}, {Coe}, {Evans}, {Waters}, \& {Jasko}}]{2018Kennea}
{Kennea}, J.~A., {Coe}, M.~J., {Evans}, P.~A., {Waters}, J., \& {Jasko}, R.~E. 2018, \apj, 868, 47, \dodoi{10.3847/1538-4357/aae839}

\bibitem[{{Lasker} {et~al.}(2008){Lasker}, {Lattanzi}, {McLean}, {Bucciarelli}, {Drimmel}, {Garcia}, {Greene}, {Guglielmetti}, {Hanley}, {Hawkins}, {Laidler}, {Loomis}, {Meakes}, {Mignani}, {Morbidelli}, {Morrison}, {Pannunzio}, {Rosenberg}, {Sarasso}, {Smart}, {Spagna}, {Sturch}, {Volpicelli}, {White}, {Wolfe}, \& {Zacchei}}]{2008Lasker}
{Lasker}, B.~M., {Lattanzi}, M.~G., {McLean}, B.~J., {et~al.} 2008, \aj, 136, 735, \dodoi{10.1088/0004-6256/136/2/735}

\bibitem[{{Maggi} {et~al.}(2013){Maggi}, {Haberl}, {Sturm}, {Pietsch}, {Rau}, {Greiner}, {Udalski}, \& {Sasaki}}]{2013Maggi}
{Maggi}, P., {Haberl}, F., {Sturm}, R., {et~al.} 2013, \aap, 554, A1, \dodoi{10.1051/0004-6361/201321238}

\bibitem[{{Maitra} {et~al.}(2023){Maitra}, {Haberl}, {Kaltenbrunner}, {Doroshenko}, {Ducci}, \& {Udalski}}]{2023Maitra}
{Maitra}, C., {Haberl}, F., {Kaltenbrunner}, D., {et~al.} 2023, The Astronomer's Telegram, 15886, 1

\bibitem[{{Martayan} {et~al.}(2007){Martayan}, {Floquet}, {Hubert}, {Guti{\'e}rrez-Soto}, {Fabregat}, {Neiner}, \& {Mekkas}}]{2007Martayan}
{Martayan}, C., {Floquet}, M., {Hubert}, A.~M., {et~al.} 2007, \aap, 472, 577, \dodoi{10.1051/0004-6361:20077390}

\bibitem[{{Mathis}(1990)}]{1990Mathis}
{Mathis}, J.~S. 1990, \araa, 28, 37, \dodoi{10.1146/annurev.aa.28.090190.000345}

\bibitem[{{Mennickent} {et~al.}(2002){Mennickent}, {Pietrzy{\'n}ski}, {Gieren}, \& {Szewczyk}}]{2002Mennickent}
{Mennickent}, R.~E., {Pietrzy{\'n}ski}, G., {Gieren}, W., \& {Szewczyk}, O. 2002, \aap, 393, 887, \dodoi{10.1051/0004-6361:20020916}

\bibitem[{{Milone} {et~al.}(2018){Milone}, {Marino}, {Di Criscienzo}, {D'Antona}, {Bedin}, {Da Costa}, {Piotto}, {Tailo}, {Dotter}, {Angeloni}, {Anderson}, {Jerjen}, {Li}, {Dupree}, {Granata}, {Lagioia}, {Mackey}, {Nardiello}, \& {Vesperini}}]{2018Milone}
{Milone}, A.~P., {Marino}, A.~F., {Di Criscienzo}, M., {et~al.} 2018, \mnras, 477, 2640, \dodoi{10.1093/mnras/sty661}

\bibitem[{{Monageng} {et~al.}(2020){Monageng}, {Coe}, {Buckley}, {McBride}, {Kennea}, {Udalski}, {Evans}, {Clark}, \& {Negueruela}}]{2020Monageng}
{Monageng}, I.~M., {Coe}, M.~J., {Buckley}, D.~A.~H., {et~al.} 2020, \mnras, 496, 3615, \dodoi{10.1093/mnras/staa1739}

\bibitem[{{Monet} {et~al.}(2003){Monet}, {Levine}, {Canzian}, {Ables}, {Bird}, {Dahn}, {Guetter}, {Harris}, {Henden}, {Leggett}, {Levison}, {Luginbuhl}, {Martini}, {Monet}, {Munn}, {Pier}, {Rhodes}, {Riepe}, {Sell}, {Stone}, {Vrba}, {Walker}, {Westerhout}, {Brucato}, {Reid}, {Schoening}, {Hartley}, {Read}, \& {Tritton}}]{2003Monet}
{Monet}, D.~G., {Levine}, S.~E., {Canzian}, B., {et~al.} 2003, \aj, 125, 984, \dodoi{10.1086/345888}

\bibitem[{{Navarete} {et~al.}(2024){Navarete}, {Ticiani dos Santos}, {Carciofi}, \& {Figueiredo}}]{2024Navarete}
{Navarete}, F., {Ticiani dos Santos}, P., {Carciofi}, A.~C., \& {Figueiredo}, A.~L. 2024, \apj, 970, 113, \dodoi{10.3847/1538-4357/ad500f}

\bibitem[{Newville {et~al.}(2024)Newville, Otten, Nelson, Stensitzki, Ingargiola, Allan, Fox, Carter, Michał, Osborn, Pustakhod, Weigand, lneuhaus, Aristov, Glenn, Mark, mgunyho, Deil, Hansen, Pasquevich, Foks, Zobrist, Frost, Stuermer, Jaskula, Caldwell, Eendebak, Pompili, Nielsen, \& Persaud}]{matt_newville_2024_10998841}
Newville, M., Otten, R., Nelson, A., {et~al.} 2024, lmfit/lmfit-py: 1.3.1, 1.3.1,  Zenodo, \dodoi{10.5281/zenodo.10998841}

\bibitem[{{Ochsenbein} {et~al.}(2000){Ochsenbein}, {Bauer}, \& {Marcout}}]{2000Ochsenbein}
{Ochsenbein}, F., {Bauer}, P., \& {Marcout}, J. 2000, \aaps, 143, 23, \dodoi{10.1051/aas:2000169}

\bibitem[{{Okazaki} \& {Negueruela}(2001)}]{2001Okazaki}
{Okazaki}, A.~T., \& {Negueruela}, I. 2001, \aap, 377, 161, \dodoi{10.1051/0004-6361:20011083}

\bibitem[{{Porter} \& {Rivinius}(2003)}]{2003Porter}
{Porter}, J.~M., \& {Rivinius}, T. 2003, \pasp, 115, 1153, \dodoi{10.1086/378307}

\bibitem[{{Reig}(2011)}]{2011Reig}
{Reig}, P. 2011, \apss, 332, 1, \dodoi{10.1007/s10509-010-0575-8}

\bibitem[{{Rezaeikh} {et~al.}(2014){Rezaeikh}, {Javadi}, {Khosroshahi}, \& {van Loon}}]{2014Rezaeikh}
{Rezaeikh}, S., {Javadi}, A., {Khosroshahi}, H., \& {van Loon}, J.~T. 2014, \mnras, 445, 2214, \dodoi{10.1093/mnras/stu1807}

\bibitem[{{Rivinius} {et~al.}(2013){Rivinius}, {Carciofi}, \& {Martayan}}]{2013Rivinius}
{Rivinius}, T., {Carciofi}, A.~C., \& {Martayan}, C. 2013, \aapr, 21, 69, \dodoi{10.1007/s00159-013-0069-0}

\bibitem[{{Roming} {et~al.}(2005){Roming}, {Kennedy}, {Mason}, {Nousek}, {Ahr}, {Bingham}, {Broos}, {Carter}, {Hancock}, {Huckle}, {Hunsberger}, {Kawakami}, {Killough}, {Koch}, {McLelland}, {Smith}, {Smith}, {Soto}, {Boyd}, {Breeveld}, {Holland}, {Ivanushkina}, {Pryzby}, {Still}, \& {Stock}}]{2005Roming}
{Roming}, P. W.~A., {Kennedy}, T.~E., {Mason}, K.~O., {et~al.} 2005, \ssr, 120, 95, \dodoi{10.1007/s11214-005-5095-4}

\bibitem[{{Rothschild} {et~al.}(2013){Rothschild}, {Markowitz}, {Hemphill}, {Caballero}, {Pottschmidt}, {K{\"u}hnel}, {Wilms}, {F{\"u}rst}, {Doroshenko}, \& {Camero-Arranz}}]{2013Rothschild}
{Rothschild}, R., {Markowitz}, A., {Hemphill}, P., {et~al.} 2013, \apj, 770, 19, \dodoi{10.1088/0004-637X/770/1/19}

\bibitem[{{Rouco Escorial} {et~al.}(2020){Rouco Escorial}, {Wijnands}, {van den Eijnden}, {Patruno}, {Degenaar}, {Parikh}, \& {Ootes}}]{2020Rouco}
{Rouco Escorial}, A., {Wijnands}, R., {van den Eijnden}, J., {et~al.} 2020, \aap, 638, A152, \dodoi{10.1051/0004-6361/201936287}

\bibitem[{{Rutledge} {et~al.}(2007){Rutledge}, {Bildsten}, {Brown}, {Chakrabarty}, {Pavlov}, \& {Zavlin}}]{2007Rutlidge}
{Rutledge}, R.~E., {Bildsten}, L., {Brown}, E.~F., {et~al.} 2007, \apj, 658, 514, \dodoi{10.1086/510183}

\bibitem[{{Schlafly} \& {Finkbeiner}(2011)}]{2011Schlafly}
{Schlafly}, E.~F., \& {Finkbeiner}, D.~P. 2011, \apj, 737, 103, \dodoi{10.1088/0004-637X/737/2/103}

\bibitem[{{Schlegel} {et~al.}(1998){Schlegel}, {Finkbeiner}, \& {Davis}}]{1998Schlegel}
{Schlegel}, D.~J., {Finkbeiner}, D.~P., \& {Davis}, M. 1998, \apj, 500, 525, \dodoi{10.1086/305772}

\bibitem[{{Schmidtke} {et~al.}(2008){Schmidtke}, {Chobanian}, \& {Cowley}}]{2008Schmidtke}
{Schmidtke}, P.~C., {Chobanian}, J.~B., \& {Cowley}, A.~P. 2008, \aj, 135, 1350, \dodoi{10.1088/0004-6256/135/4/1350}

\bibitem[{{Skowron} {et~al.}(2021){Skowron}, {Skowron}, {Udalski}, {Szyma{\'n}ski}, {Soszy{\'n}ski}, {Wyrzykowski}, {Ulaczyk}, {Poleski}, {Koz{\l}owski}, {Pietrukowicz}, {Mr{\'o}z}, {Rybicki}, {Iwanek}, {Wrona}, \& {Gromadzki}}]{2021Skowron}
{Skowron}, D.~M., {Skowron}, J., {Udalski}, A., {et~al.} 2021, \apjs, 252, 23, \dodoi{10.3847/1538-4365/abcb81}

\bibitem[{{Stella} {et~al.}(1986){Stella}, {White}, \& {Rosner}}]{1986Stella}
{Stella}, L., {White}, N.~E., \& {Rosner}, R. 1986, \apj, 308, 669, \dodoi{10.1086/164538}

\bibitem[{STScI(2020)}]{2020STScI}
STScI. 2020, Digitized Sky Survey,  IPAC, \dodoi{10.26131/IRSA441}

\bibitem[{{Sturm} {et~al.}(2013){Sturm}, {Haberl}, {Pietsch}, {Ballet}, {Hatzidimitriou}, {Buckley}, {Coe}, {Ehle}, {Filipovi{\'c}}, {La Palombara}, \& {Tiengo}}]{2013Sturm}
{Sturm}, R., {Haberl}, F., {Pietsch}, W., {et~al.} 2013, \aap, 558, A3, \dodoi{10.1051/0004-6361/201219935}

\bibitem[{{Townsend} {et~al.}(2017){Townsend}, {Kennea}, {Coe}, {McBride}, {Buckley}, {Evans}, \& {Udalski}}]{2017Townsend}
{Townsend}, L.~J., {Kennea}, J.~A., {Coe}, M.~J., {et~al.} 2017, \mnras, 471, 3878, \dodoi{10.1093/mnras/stx1865}

\bibitem[{{Treiber} {et~al.}(2025){Treiber}, {Vasilopoulos}, {Bailyn}, {Haberl}, \& {Udalski}}]{2025Treiber}
{Treiber}, H., {Vasilopoulos}, G., {Bailyn}, C.~D., {Haberl}, F., \& {Udalski}, A. 2025, \aap, 694, A43, \dodoi{10.1051/0004-6361/202451132}

\bibitem[{{Udalski} {et~al.}(1997){Udalski}, {Kubiak}, \& {Szymanski}}]{1997Udalski}
{Udalski}, A., {Kubiak}, M., \& {Szymanski}, M. 1997, \actaa, 47, 319, \dodoi{10.48550/arXiv.astro-ph/9710091}

\bibitem[{{Wenger} {et~al.}(2000){Wenger}, {Ochsenbein}, {Egret}, {Dubois}, {Bonnarel}, {Borde}, {Genova}, {Jasniewicz}, {Lalo{\"e}}, {Lesteven}, \& {Monier}}]{2000Wenger}
{Wenger}, M., {Ochsenbein}, F., {Egret}, D., {et~al.} 2000, \aaps, 143, 9, \dodoi{10.1051/aas:2000332}

\bibitem[{{Yang} {et~al.}(2017){Yang}, {Laycock}, {Christodoulou}, {Fingerman}, {Coe}, \& {Drake}}]{2017Yang}
{Yang}, J., {Laycock}, S.~G.~T., {Christodoulou}, D.~M., {et~al.} 2017, \apj, 839, 119, \dodoi{10.3847/1538-4357/aa6898}

\bibitem[{{Yokogawa} {et~al.}(2000){Yokogawa}, {Imanishi}, {Tsujimoto}, {Nishiuchi}, {Koyama}, {Nagase}, \& {Corbet}}]{2000Yokogawa}
{Yokogawa}, J., {Imanishi}, K., {Tsujimoto}, M., {et~al.} 2000, \apjs, 128, 491, \dodoi{10.1086/313394}

\bibitem[{{Zacharias} {et~al.}(2004){Zacharias}, {Monet}, {Levine}, {Urban}, {Gaume}, \& {Wycoff}}]{2004Zacharias}
{Zacharias}, N., {Monet}, D.~G., {Levine}, S.~E., {et~al.} 2004, in American Astronomical Society Meeting Abstracts, Vol. 205, American Astronomical Society Meeting Abstracts, 48.15

\bibitem[{{Zezas} {et~al.}(2003){Zezas}, {McDowell}, {Hadzidimitriou}, {Kalogera}, {Fabbiano}, \& {Taylor}}]{2003Zezas}
{Zezas}, A., {McDowell}, J.~C., {Hadzidimitriou}, D., {et~al.} 2003, in The Local Group as an Astrophysical Laboratory, ed. M.~{Livio} \& T.~M. {Brown}, 111, \dodoi{10.48550/arXiv.astro-ph/0310562}

\end{thebibliography}
\bibliographystyle{aasjournal}



\end{document}